\documentclass[12pt]{elsarticle}

\usepackage[margin=1.1in]{geometry}

\usepackage{amsmath, amssymb}
\usepackage{bm}
\usepackage{mathrsfs}
\usepackage{graphicx, fancyhdr}
\usepackage{placeins} 
\usepackage{floatrow}

\usepackage{ifpdf}
\usepackage{multirow}
\usepackage{caption}
\usepackage{subcaption}

\newdimen\figrasterwd
\figrasterwd\textwidth

\usepackage{setspace}
\usepackage{xcolor}

\usepackage{hyperref}
\hypersetup{
 colorlinks=true,
 linkcolor=blue,
 citecolor=blue,
 urlcolor=blue
 }

\journal{Journal of Computational Physics}
\begin{document}

\begin{frontmatter}

\title{Graph Network Surrogate Model for Subsurface Flow Optimization}

\author{Haoyu Tang\corref{mycorrespondingauthor}}
\ead{hytang@stanford.edu}

\author{Louis J.~Durlofsky}
\ead{lou@stanford.edu}
\address{Department of Energy Science \& Engineering, Stanford University, Stanford, CA 94305, USA}

\cortext[mycorrespondingauthor]{Corresponding author}

\begin{abstract}
The optimization of well locations and controls is an important step in the design of subsurface flow operations such as oil production or geological CO$_2$ storage. These optimization problems can be computationally expensive, however, as many potential candidate solutions must be evaluated. In this study, we propose a graph network surrogate model (GNSM) for optimizing well placement and controls. The GNSM transforms the flow model into a computational graph that involves an encoding-processing-decoding architecture. Separate networks are constructed to provide global predictions for the pressure and saturation state variables. Model performance is enhanced through the inclusion of the single-phase steady-state pressure solution as a feature. A multistage multistep strategy is used for training. The trained GNSM is applied to predict flow responses in a 2D unstructured model of a channelized reservoir. Results are presented for a large set of test cases, in which five injection wells and five production wells are placed randomly throughout the model, with a random control variable (bottom-hole pressure) assigned to each well. Median relative error in pressure and saturation for 300 such test cases is 1-2$\%$. The ability of the trained GNSM to provide accurate predictions for new (geologically similar) permeability realizations
is demonstrated. Finally, the trained GNSM is used to optimize well locations and controls with a differential evolution algorithm. GNSM-based optimization results are comparable to those from simulation-based optimization, with a runtime speedup factor of 36. Much larger speedups are expected if the method is used for robust optimization, in which each candidate solution is evaluated on multiple geological models.

\end{abstract}


\end{frontmatter}


\section{Introduction}
The efficient design and operation of subsurface flow processes, such as oil/gas and geothermal energy production, geological CO$_2$ storage, and freshwater aquifer management, require the solution of a variety of optimization problems. A primary issue in many settings is the determination of optimal well locations. This well placement optimization problem provides the motivation for the methodology developed in this study. Well placement problems typically display multiple local optima, so gradient-based methods can encounter challenges. Metaheuristic global stochastic search algorithms are, therefore, often applied for these optimizations, though these methods are computationally expensive due to the large number of function evaluations (multiphase flow simulations) required. 
In this study, we develop and apply a graph network surrogate model that can estimate flow results in 2D unstructured models with varying well configurations and well controls.

Machine learning (ML) and deep learning (DL) techniques are now widely applied for subsurface flow problems. Application areas outside of well placement optimization include well control optimization and data assimilation (history matching). Well control optimization involves determining injection and production rates or bottom-hole pressures (BHPs) to maximize a performance metric. Machine learning algorithms, such as artificial neural networks (ANNs)~\citep{chai2021integrated}, recurrent neural networks (RNNs)~\citep{kim2023convolutional}, convolutional neural networks (CNNs)~\citep{kim2023convolutional}, and reinforcement learning \citep{nasir2023deep}, are a few of the methods that have been used for these problems. Data assimilation entails the use of observed (historical) data to reduce uncertainty in model parameters such as grid-block values of permeability and porosity. Many investigators have utilized deep-learning procedures for history matching. The approaches used in this context include theory-guided neural networks (TgNNs) \cite{xu2022uncertainty,wang2022surrogate}, the CNN-based recurrent residual U-Net \citep{tang2021deep}, physics-informed neural networks (PINNs) \citep{geneva2020modeling}, and generative adversarial networks (GANs)~\citep{razak2022conditioning}. Deep-learning-based methods are also now used in geological CO$_2$ storage settings, e.g., \cite{tang2022deep,wen2022accelerating,grady2022towards}.

In the context of well placement optimization, many current DL or ML-based surrogate models focus on the direct prediction of the objective function, e.g., cumulative oil production or net present value (NPV). Kim et al.~\citep{kim2020robust}, for example, developed a CNN-based approach that used time of flight maps (derived from streamline simulations) for each well configuration to predict NPV for two-phase oil-water problems. Tang and Durlofsky~\citep{tang2022use} applied tree-based machine learning models (random forest and gradient boosting) to correct error in NPV due to the use of low-fidelity (upscaled) geomodels for the 3D oil-water flow simulations.
Nwachukwu et al.~\citep{nwachukwu2018fast} and Mousavi et al.~\citep{mousavi2020optimal} utilized tree-based machine learning methods (XGBoost) to predict NPV for well placement optimization problems. These studies considered waterflood and CO$_2$ injection~\citep{nwachukwu2018fast}, as well as oil-gas systems \citep{mousavi2020optimal}. 
Redouane et al.~\citep{redouane2018automated} and Bruyelle and Gu{\'e}rillot~\citep{bruyelle2019well} applied artificial neural network (ANN) surrogates to predict cumulative oil and NPV as a function of well locations in 3D waterflood models. Wang et al.~\citep{wang2022efficient} applied a TgNN-based surrogate model to predict pressure state maps, which they used to optimize production-well locations in 2D single-phase-flow problems. Although there has been significant progress in developing DL/ML-based surrogate models for well placement optimization, further progress is needed to provide models that can predict more complete state information for multiphase flow problems in complex geological settings.

Graph neural network (GNN) models are a class of deep-learning methods designed to handle data in terms of nodes and edges. Many physical problems contain particles and/or cells that can be represented as nodes, while their interactions can be treated as edges. Because GNN models predict the interactions between nodes and edges, they are suitable for many science and engineering problems. Example GNN applications include simulating the motion of complex solid objects~\citep{sanchez2020learning, pfaff2020learning,wulearning}, fluid flow simulation~\citep{zhao2022learning, li2022graph, belbute2020combining}, and weather forecasting~\citep{lienen2022learning,lam2022graphcast}. 

In the context of subsurface flow simulation, Wu et al.~\cite{wu2022learning} developed a hybrid deep-learning surrogate model that combined a 3D U-Net for predicting pressure with a GNN model for saturation. A 3D oil-water problem was considered in this study. Different static properties, initial states, and well locations were used for training and testing. Well locations were varied in this work by activating and deactivating a predetermined set of wells with fixed locations -- arbitrary well configurations were not considered. Nonetheless, this work demonstrated the potential of GNN-based surrogate models for predicting state variables in large 3D reservoir models. Further development and application of this approach for well placement optimization is clearly warranted.

In this study, we present a graph network surrogate model, referred to as GNSM, for use in well placement optimization problems involving 2D unstructured models. Graph networks (GNs) are a type of GNN that possess an encoding-processing-decoding structure. This structure is built on a computational graph in which 
the relationships and interactions between nodes in the graph are predicted. 
Our GNSM employs separate GNs, with different hyperparameter settings, to predict time-varying pressure and saturation. 
This model differs from that in Wu et al.~\cite{wu2022learning} in that we use GNs as the surrogate model for both pressure and saturation predictions. In addition to allowing wells to be in any grid cell (and not just in a set of predetermined locations), we also treat the well controls, specifically BHP for each well, as variables. 
The GNSM is trained using multistage training strategies for well configurations and controls relevant to the target optimization problem. In particular, we consider 10-well configurations (five injectors and five producers) in unstructured 2D models of oil-water systems. During the testing stage, the GNSM is applied to cases with new well configurations and BHP settings. A differential evolution (DE) algorithm is applied for the optimizations, and GNSM-based optimization results are compared to those from simulation-based optimization.

The paper proceeds as follows. In Section~\ref{sec:equ_and_opt}, we present the governing equations for the oil-water system and describe the well placement optimization problem. The GNSM is introduced in Section~\ref{sec:gnsm}, where we discuss model architecture, the key input features and hyperparameters, and training strategies. The use of the GNSM for well placement optimization is also explained. Detailed test-case results for an unstructured-grid model are provided in Section~\ref{sec:result_test}, and optimization results appear in Section~\ref{sec:result_opt}. We summarize our findings and provide suggestions for future work in Section~\ref{sec:summary}. In Appendix~A, we present summary results for structured-grid cases. 

\section{Governing equations and optimization procedure}
\label{sec:equ_and_opt}
In this section, we present the governing flow equations and describe the optimization problem considered in this work.

\subsection{Flow equations}
\label{sec:gov_equations}

We consider immiscible two-phase subsurface flow in 2D systems. This formulation is used for oil reservoir simulation as well as for other applications such as aquifer remediation. The fluid phases are denoted by subscript $j$, where $j=o$ indicates oil and $j=w$ water. Mass conservation for each component (which exists only in its own phase) can be expressed as:

\begin{flalign}  \label{eq:fluid mass balance}
    \frac{\partial}{\partial{t}}\left(\phi \rho_j S_j \right) + \nabla \cdot \left( \rho_j \mathbf{v}_j \right) = -\tilde{q}^m_j,
\end{flalign}

\noindent where $t$ is time, $\phi$ is porosity, $\rho_j$ is phase density, $S_j$ is  phase saturation, $\mathbf{v}_j$ is the Darcy phase velocity, and $\tilde{q}_j^m$ indicates the mass source/sink term (superscript $m$ denotes mass). In many subsurface flow problems, the source terms are driven by injection and production wells ($\tilde{q}_j^m>0$ for production). Here, tilde indicates the source terms are per unit volume, meaning $\tilde{q}^m_j$ is of dimensions mass/(volume-time). Darcy velocity for phase $j$ is given by

\begin{equation} 
     \mathbf{v}_j =  -\frac{\mathbf{k} k_{rj}(S_j)}{\mu_j (p_j)} \nabla p_j,
\label{eq:darcy-eq}
\end{equation}
\noindent where $\mathbf{k}$ is the permeability tensor, $k_{rj}$ denotes the relative permeability for phase $j$, $\mu_j$ is the phase viscosity, and $p_j$ is the phase pressure. As is common in large-scale reservoir simulation, we take $p_o=p_w$, meaning capillary pressure effects are neglected.

In our GNSM, we will construct separate networks for the pressure and saturation variables. To motivate this treatment, we reformulate Eqs.~\ref{eq:fluid mass balance} and \ref{eq:darcy-eq} into so-called pressure and saturation equations. Under the assumption of incompressible fluid and rock, applicable (at least approximately) for many two-phase systems, the pressure equation is given by:
\begin{equation} \label{eq:pressure equation}
\nabla \cdot \left[\lambda_t(S_w) \mathbf{k} \nabla p \right] = \tilde{q}^v_t,  
\end{equation}
\noindent where $\lambda_t(S_w) = k_{rw}/\mu_w + k_{ro}/\mu_o$ is the total mobility and $\tilde{q}^v_t = \tilde{q}^m_w/\rho_w + \tilde{q}^m_o/\rho_o$ is the total volumetric source term (dimensions of 1/time), where superscript $v$ indicates volumetric. The water saturation equation is expressed as:
\begin{equation} \label{eq:saturation equation}
\phi \frac{\partial S_w}{\partial{t}} + \nabla \cdot \left[ {\mathbf v}_t f(S_w) \right] = -\tilde{q}^v_w,
\end{equation}
\noindent where ${\mathbf v}_t={\mathbf v}_w+{\mathbf v}_o = -\lambda_t(S_w) \mathbf{k}  \nabla p$ is the total velocity, $f(S_w)=k_{rw}/(\mu_w \lambda_t)$ is the flux function and $\tilde{q}^v_w = \tilde{q}^m_w/\rho_w$. From this representation, we see that pressure is governed by an elliptic partial differential equation, while the saturation equation is hyperbolic. 

In reservoir simulation, the governing equations (either Eqs.~\ref{eq:fluid mass balance} and \ref{eq:darcy-eq} or Eqs.~\ref{eq:pressure equation} and \ref{eq:saturation equation}) are discretized and solved using finite volume procedures. When we integrate over the well block, the source terms, now denoted $q^v_w$ and $q^v_o$, are of dimensions volume/time. These terms are modeled using the Peaceman~\citep{peaceman1983interpretation} representation, given by:

\begin{equation} 
     (q^v_j)_i = W_i\left(\frac{k_{rj}}{\mu_j} \right)_i (p_i-p^w),
\label{eq:well_rate-eq}
\end{equation}
where $i$ denotes the finite volume cell, $(q^v_j)_i$ is the volumetric flow rate of phase $j$ into or out of the well in grid block $i$, and $W_i$ is the well index. The latter quantity depends on the grid block and wellbore geometry, and rock properties, and is given by
\begin{equation} 
     W_i = \frac{2 \pi k_i \Delta z}{\ln \left(\frac{r_w}{r_0}\right)}.
\label{eq:wellindex-eq}
\end{equation}
\noindent Here $k_i$ is the cell permeability, $\Delta z$ is the cell thickness, $p_i$ and $p^w$ represent the well-cell pressure and wellbore pressure (which is the same as BHP in our 2D model), and $r_w$ is the wellbore radius. For the unstructured models considered here, $r_0 = 0.2(V/\Delta z)^{1/2}$, where $V$ is the bulk volume of the cell~\citep{lie2019introduction}. This representation, which assumes permeability is isotropic, reduces to the usual expression, $r_0 = 0.2 \Delta x$, for square grid blocks. 
In the simulations performed here, we specify wellbore pressure/BHP, $p^w$. Eqs.~\ref{eq:well_rate-eq} and \ref{eq:wellindex-eq} are then applied to obtain the well volumetric flow rates, which are key quantities of interest in many simulation studies.

Flow simulations in this work are performed using Stanford's Automatic Differentiation General Purpose Research Simulator, ADGPRS~\citep{zhou2012parallel}. These simulation results provide training data as well as the reference against which we will compare GNSM predictions.

\subsection{Optimization formulation and algorithm}
\label{sec:opt_algo}

The optimization problem considered here entails the determination of the locations of a specified number of injection and production wells, along with their corresponding well controls, such that an objective function is maximized or minimized. The goal in this work is to maximize the net present value (NPV) of the project over a specified time frame. The optimization problem is stated as

\begin{align}
\begin{cases}
\underset{\mathbf{u}\in \mathbb{U}}{\max} \;\; \mathit{J}(\mathbf{u}),  \\
\mathbf{c}(\mathbf{u}) \leq \mathbf{0}.  \label{eqn:wpo_def}
\end{cases}
\end{align}

\noindent Here $J$ is the objective function we seek to maximize, $\mathbf{u}\in \mathbb{U} \subset{\mathbb{R}^{3n_w}}$ contains the $x$ and $y$ coordinates of each well plus the constant-in-time BHP for each well (there are a total of $n_w$ wells). The $x$-$y$ locations within the unstructured grid and the well controls must lie within the feasible region $\mathbb{U}$. The vector $\mathbf{c}$ represents any nonlinear constraints that are imposed. Such constraints can involve the well geometry and/or simulation output quantities (e.g., minimum oil rate, maximum water fraction). NPV involves discounted revenue from oil production and costs for water injection and production. The detailed expression will be given in Section~\ref{sec:application}. 

\textcolor{black}{With NPV as the objective function, only well-rate quantities are involved in the optimization. Other objective functions, though not considered in this paper, will require global state information. Examples include maximizing sweep or minimizing bypassed oil in reservoir simulation, and maximizing storage efficiency or minimizing mobile CO$_2$ in carbon storage modeling.}

Optimization is accomplished in this study using differential evolution (DE). The general DE algorithm is described by Price et al.~\citep{price2006differential}. The method was used for well placement optimization by Zou et al.~\citep{zou2022effective}, who found it to outperform some of the other algorithms that are commonly used for this application. DE is a population-based global stochastic search method. It resembles genetic algorithms (GAs) in that both approaches contain mutation, crossover, and selection processes. Unlike many GAs, DE uses real numbers to represent the optimization variables instead of a binary representation. More detailed discussion on the DE implementation used in this study can be found in~\citep{tang2022use}.

\section{GNSM framework and applications}
\label{sec:gnsm}

Our goal is to predict state variables (pressure and saturation) in all grid blocks in the model at a set of time steps. As noted earlier, in an incompressible system the pressure equation (Eq.~\ref{eq:pressure equation}) is elliptic and the saturation equation (Eq.~\ref{eq:saturation equation}) is hyperbolic. We therefore develop two GNNs -- a pressure graph neural network (PresGNN) and a saturation graph neural network (SatGNN) -- with similar architectures but different hyperparameters. These GNNs are trained offline independently. During the online inference stage, input (at time step $n$) is provided to both GNNs, and the outputs are pressure and saturation at the next time step ($n+1$). These results then provide input for the computations at time step $n+2$, etc. This process, referred to as one-step rollout, is analogous to the first-order time stepping typically applied in reservoir simulation. We now describe the surrogate model architecture, input features and hyperparameters, training strategies, and the online application procedure.

\subsection{Model architecture}
\label{sec:model}

\begin{figure*}[!htb]
\centering
\includegraphics[width = 1.0\textwidth]{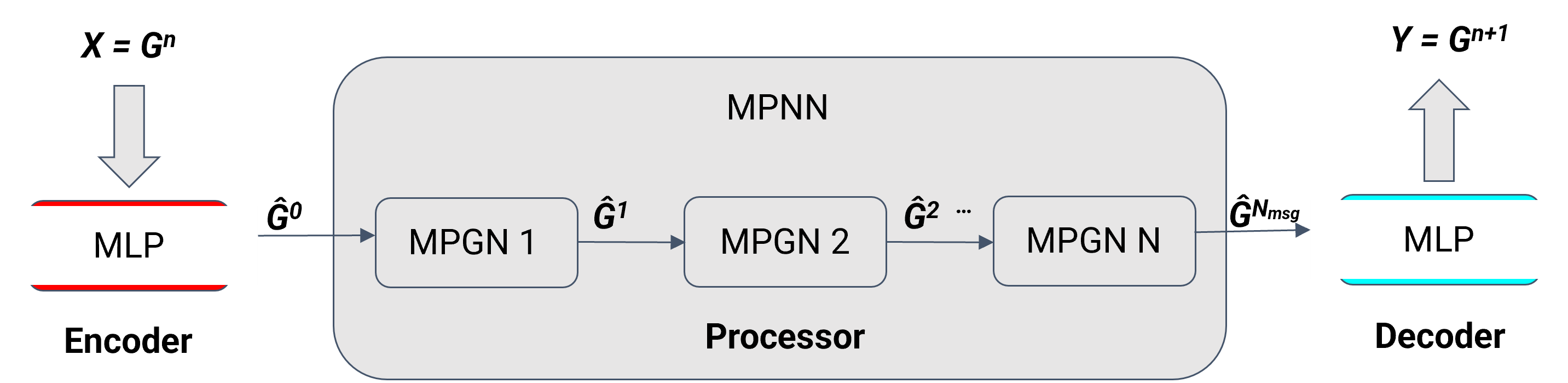}
\caption{Schematic of model architecture (encoder-processor-decoder). The encoder and decoder are MLPs. The processor is a message passing neural network containing $N_{msg}$ message passing graph networks (MPGNs).} \label{fig:model_architecture}
\end{figure*}

The basic architecture for both the pressure and saturation GNNs is illustrated in Fig.~\ref{fig:model_architecture}. This architecture involves an encoder, a processor, and a decoder. This design shares similarities with both the graph network-based simulator (GNS) in~\citep{sanchez2020learning} and the subsurface graph neural network (SGNN) in~\citep{wu2022learning}. There are some important differences, as 
discussed in Section~\ref{sec:features_and_hyper}, which act to improve performance and robustness.

The encoder performs a global encoding for each node and edge of the graph input $\mathbf G^n = \left(\mathbf N^n, \mathbf E\right)$. This results in the latent graph $\hat {\mathbf G}^0 = \left(\hat{\mathbf N}^0, \hat{\mathbf E}^0\right)$. Here, $\mathbf G^n$ refers to the input graph at time step $n$, $\mathbf N^n$ denotes the node features at time step $n$, $\mathbf E$ contains the edge features, and $\hat {\mathbf G}^0$ is the encoded graph. The encoder contains a multilayer perceptron (MLP). 

The processor is a message-passing neural network (MPNN) that contains $N_{msg}$ message-passing graph networks (MPGNs)~\citep{sanchez2020learning}, where $N_{msg}$ is the number of message-passing layers (specified as a hyperparameter). The value of $N_{msg}$ determines the set of neighbors from which the GNN collects information. As we will see, the optimal value of this hyperparameter differs between PresGNN and SatGNN. For each MPGN $i$, $i=1,\dots,N_{msg}$, the input is denoted $\mathbf G^{i-1}$ (from the previous MPGN) and the output $\mathbf G^{i}$.

\begin{figure*}[!htb]
\centering
\includegraphics[width = 0.9\textwidth]{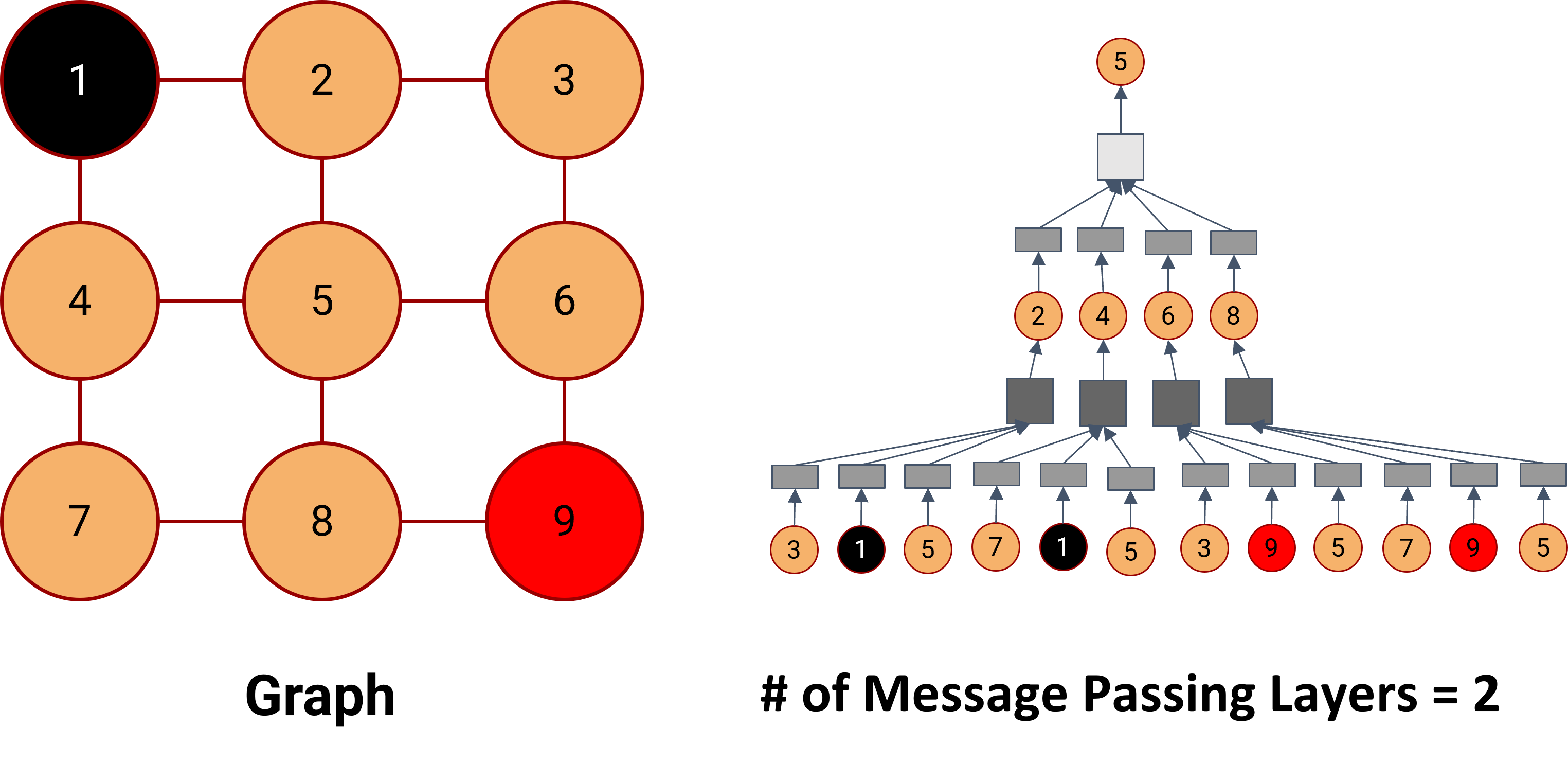}
\caption{MPNN schematic for $3 \times 3$ reservoir model (left). In the computational graph shown on the right, Node~5 is the target node to be updated with $N_{msg}=2$. Node~1 (black) corresponds to a cell with an injection well, and Node~9 (red) to a cell with a production well.} \label{fig:message_passing}
\end{figure*}

To simplify the presentation, we first describe the MPNN for a structured model. Fig.~\ref{fig:message_passing} shows a schematic of an MPNN for a reservoir model containing $3 \times 3$ cells with $N_{msg} = 2$. The reservoir cells are represented by Nodes~1--9 in the left image, and the lines connecting them represent edges. Node~1 (black) indicates a cell with an injection well, and Node~9 (red) is a cell with a production well. A computational graph is generated for each node to be updated. The image on the right shows the computational graph with Node~5 as the target node. We use the term `matrix node' to refer to a cell/node that does not contain a well (thus Nodes~2--8 are matrix nodes).  

Given that we have specified $N_{msg} = 2$ for this case, information is communicated across two layers, as shown in Fig.~\ref{fig:message_passing} (right). The rectangular boxes represent information collection, and the square boxes indicate information aggregation. As indicated, the network updates the state for Node~5 using information from Nodes~1--9. The final encoded graph after $N_{msg}$ steps of the MPGN (applied for every node) is denoted $\hat {\mathbf G}^{N_{msg}}$.

\begin{figure*}[!htb]
\centering
\includegraphics[width = 0.6\textwidth]{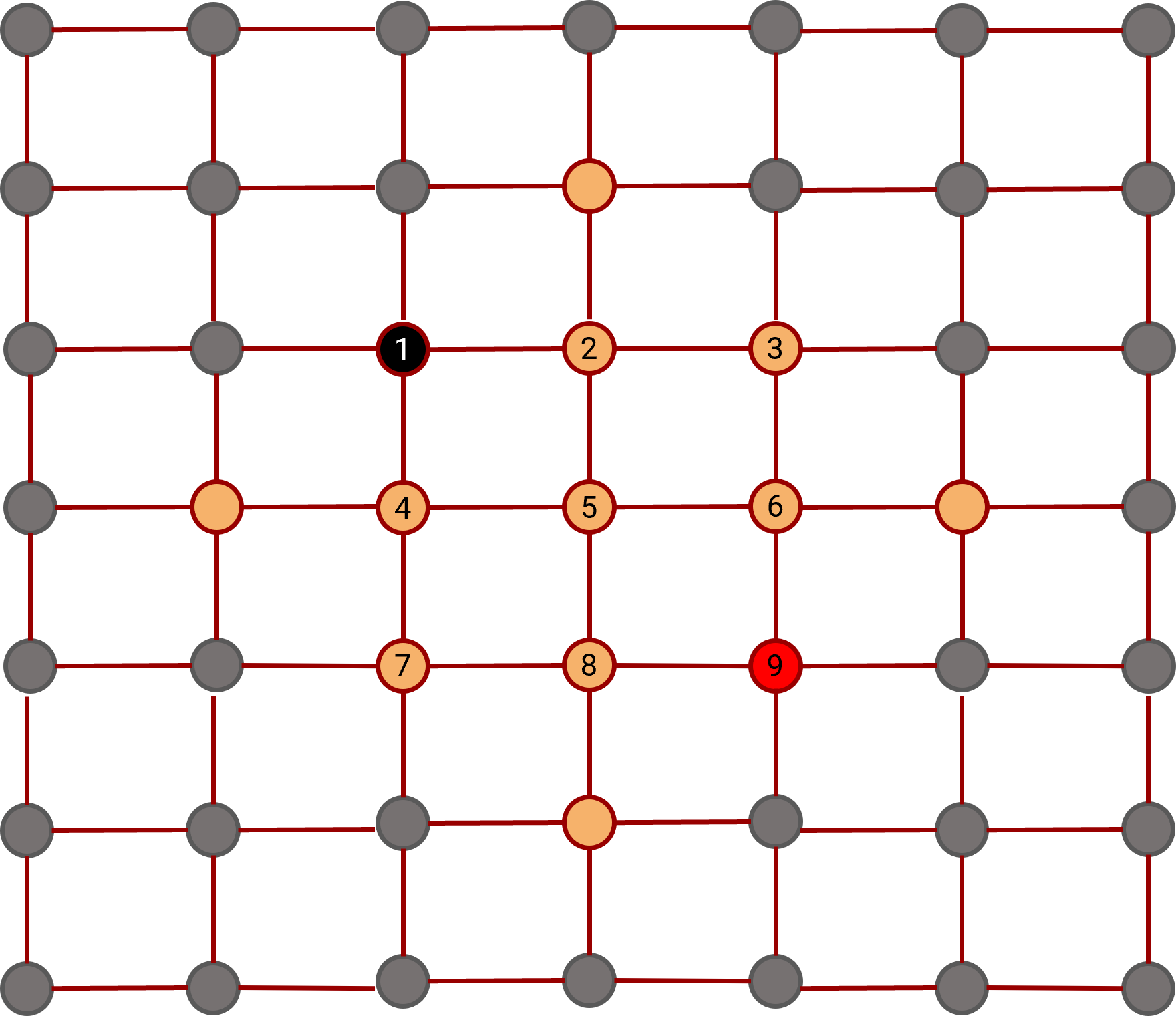}
\caption{MPNN example for a $7 \times 7$ model, with Node~5 as the target node. Nodes containing injection and production wells shown in black and red. Gray and brown nodes are matrix nodes. Gray nodes do not contribute to the computational graph for Node~5 (with $N_{msg} = 2$).} \label{fig:big_cell_example_mpnn}
\end{figure*}

All nodes appear in the computational graph in this simple structured example. In realistic models, however, most nodes/cells will be outside the range of the $N_{msg}$ layers (unless $N_{msg}$ is specified to be very large). This is illustrated in Fig.~\ref{fig:big_cell_example_mpnn}, where we indicate the nodes that directly contribute to the computational graph for a larger ($7 \times 7$) system, again with $N_{msg}=2$. In this case, the gray nodes do not appear in the computational graph for (target) Node~5. Thus, we see that the number of connections will be limited, even in large models, through use of reasonable $N_{msg}$ values. 

The approach with unstructured models is analogous to that depicted in Figs.~\ref{fig:message_passing} and \ref{fig:big_cell_example_mpnn}. In the unstructured case, the computational graph shown in Fig.~\ref{fig:message_passing} (right), for $N_{msg}=2$, is modified to include the neighbors of the target cell (in the second row) and the neighbors of those cells (in the third row). The numbers of cells in these rows will, in general, differ for each target cell. Information collection and aggregation (gray rectangles and squares in Fig.~\ref{fig:message_passing}, right) is modified accordingly. The network structure is otherwise the same as that described above.

The decoder (Fig.~\ref{fig:model_architecture}) performs a process that is the reverse of that in the encoder. The processed graph data $\hat {\mathbf G}^N$ are passed to the decoding MLP. This provides $\mathbf G^N$, which contains the updated state variables (pressure or saturation) for the current time step $n$.

\subsection{Features and hyperparameters}
\label{sec:features_and_hyper}

\begin{figure*}[!htb]
\centering
\begin{subfigure}{.285\linewidth}\centering
\includegraphics[width=\linewidth]{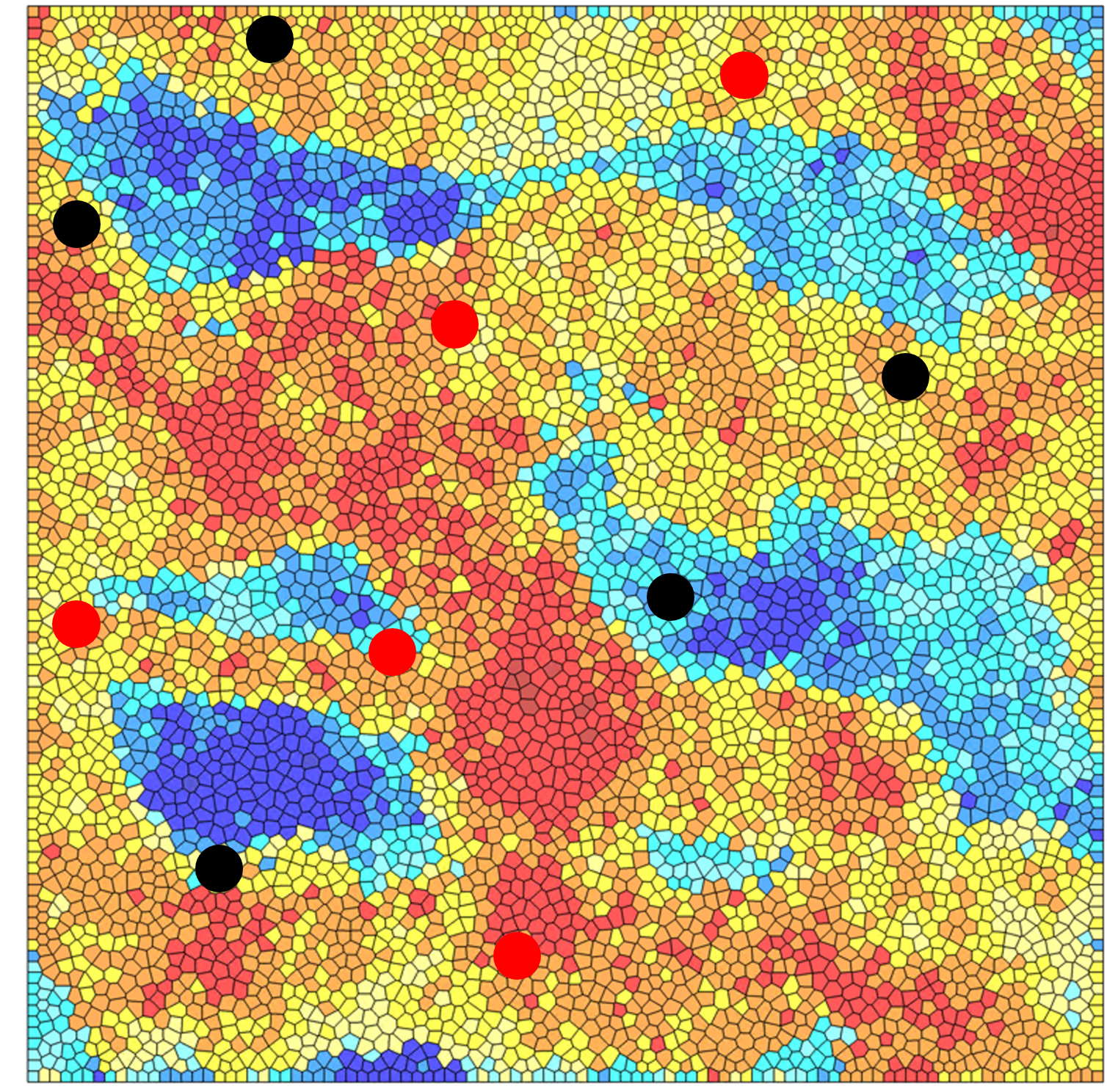}\caption{Well Configuration 1}
\end{subfigure}
\begin{subfigure}{.285\linewidth}\centering
\includegraphics[width=\linewidth]{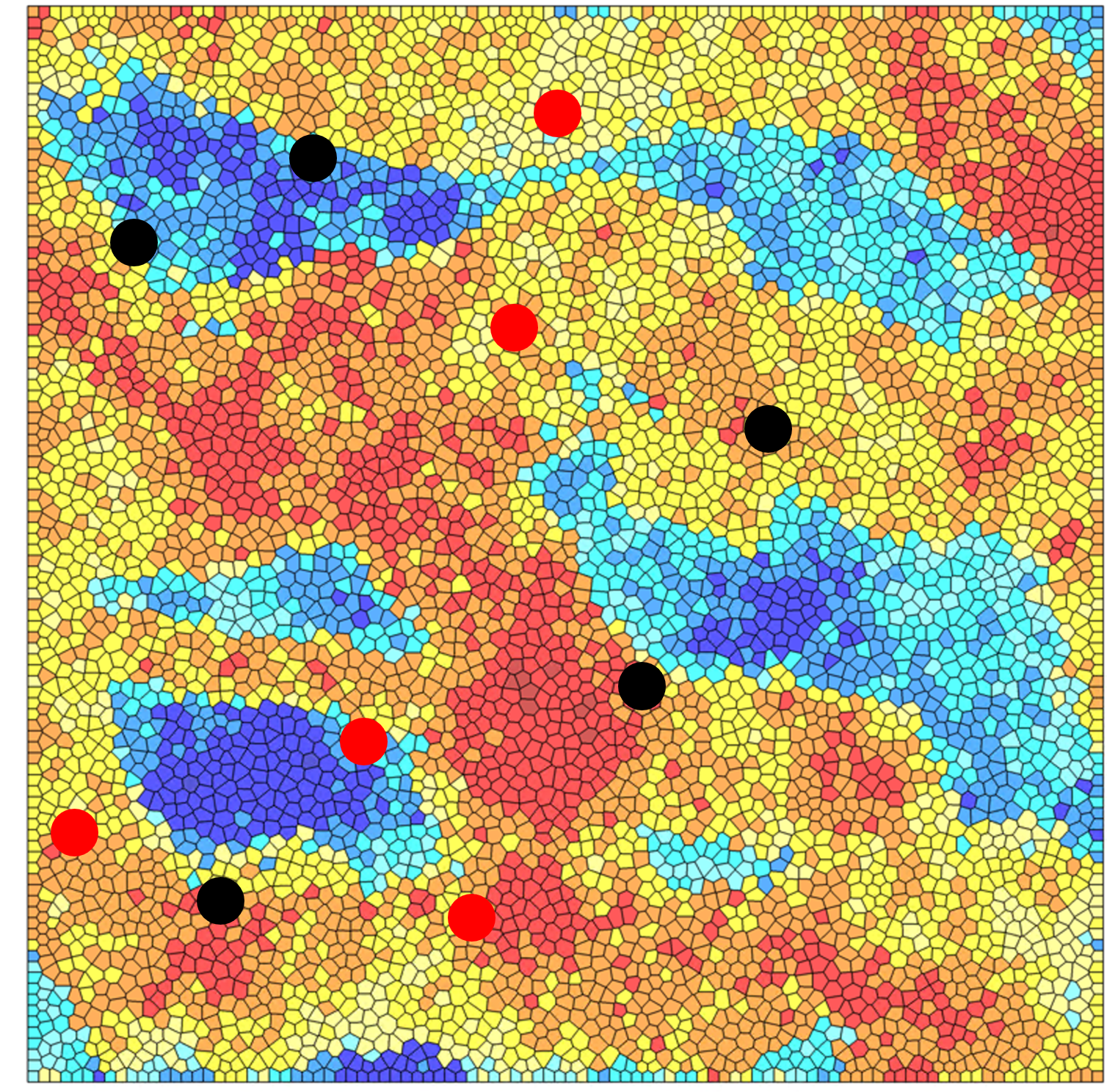}\caption{Well Configuration 2}
\end{subfigure}
\begin{subfigure}{.358\linewidth}\centering
\includegraphics[width=\linewidth]{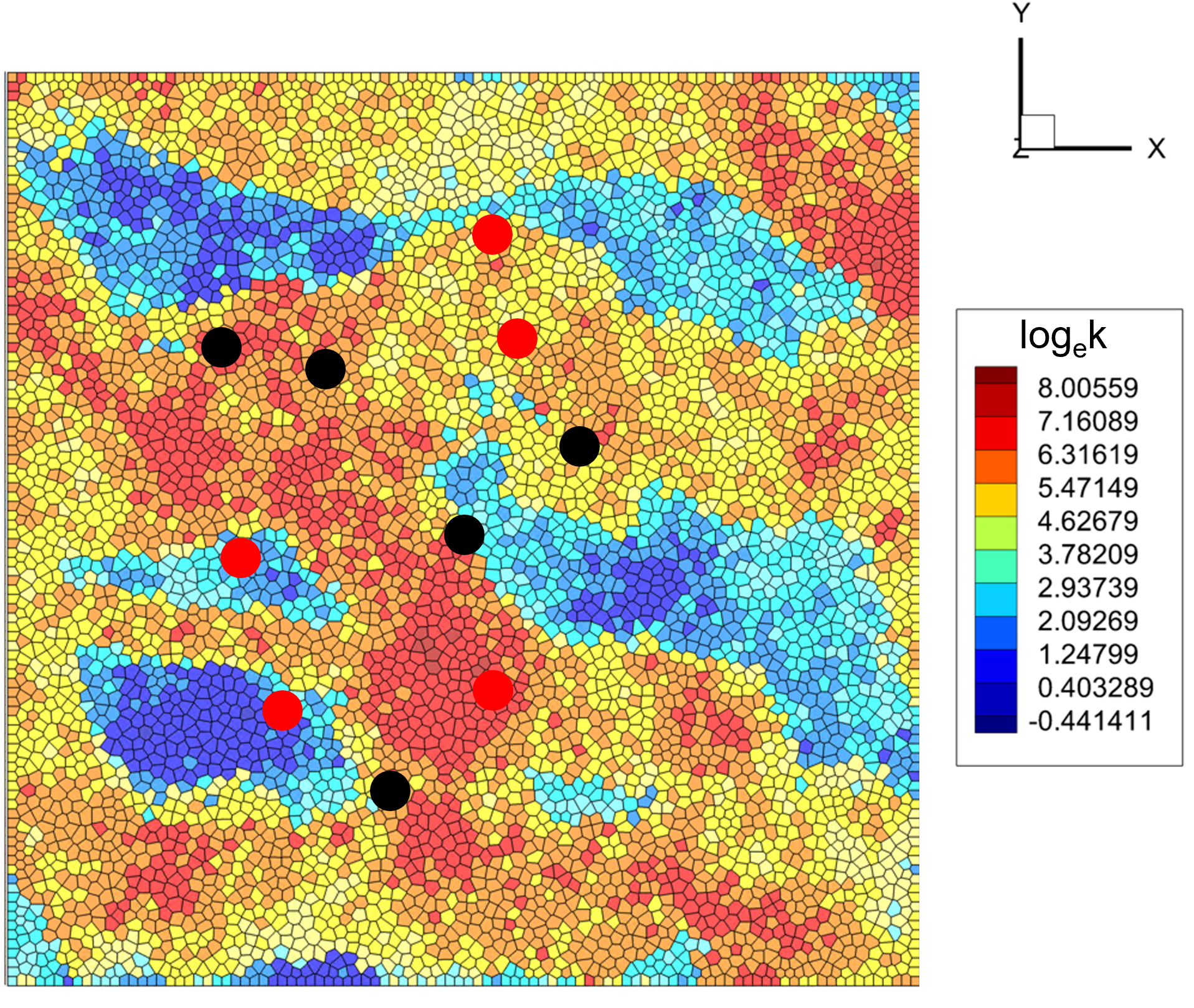}\caption{Well Configuration 3}
\end{subfigure}
\caption{Three example well configurations. All training configurations contain five injection wells (black circles) and five production wells (red circles). The background shows the log-permeability field (in md) and the unstructured simulation grid.} \label{fig:setup_example}
\end{figure*}

To motivate the features used in the GNSM, we first describe the general problem setup. We consider a 2D unstructured reservoir model, with oil produced via waterflood. There are five water injection wells and five production wells, with each well operating at a different constant-in-time BHP ($p^w$ in Eq.~\ref{eq:well_rate-eq}) over the simulation time frame. The geological model, which corresponds to a channelized system, is fixed across all training runs. In each training run, the wells are distributed randomly throughout the model (configurations in which any two wells are less than 100~m apart are discarded). Three example configurations are shown in Fig.~\ref{fig:setup_example}. The background shows the permeability field ($\log_e k$, with $k$ in md, is displayed). The well configurations differ significantly between the three examples. The BHP for each injection well is randomly sampled from the uniform distribution [210, 310]~bar. Production well BHPs are sampled from [50, 150]~bar.
The wide range of configurations and BHPs applied during training give rise to very different oil/water production profiles.

The GNSM works with graph data involving nodes and edges, so the GNSM features are divided into node and edge features. These two sets of features are listed in Tables~\ref{tab:node_features} and \ref{tab:edge_features}. Note that feature~6 in Table~\ref{tab:node_features} and feature~4 in Table~\ref{tab:edge_features} are not relevant in the 2D horizontal models considered here, but these are required for more general cases (3D or 2D models with dip).

In Table~\ref{tab:node_features}, features~1 and 2 are state variables for the node at time~$n$. Features~3--6 are geological and geometric properties of the node -- specifically isotropic permeability, porosity, bulk volume ($V$), and depth of the target cell. Features~7 and 8 relate to wells -- for matrix nodes, they are both set to 0. The well index, $W$, is defined in Eq.~\ref{eq:wellindex-eq}. Feature~9, the encoding ${e}$, defines the node type, with $[1, 0 ,0]$ denoting a cell containing an injection well, $[0, 1, 0]$ a cell containing a production well, and $[0, 0, 1]$ a matrix cell. 

The last feature in Table~\ref{tab:node_features} is the pressure field for a single-phase steady-state simulation. This is a separate solution that must be computed during training and in subsequent online calculations. We have found this extra computation to be worthwhile, as it leads to improved GNSM accuracy. In this solution, we set $\lambda_t=1/\mu_1$ in Eq.~\ref{eq:pressure equation}, where $\mu_1$ is the constant (arbitrary) single-phase viscosity, which gives
\begin{equation} \label{eq:1p_pressure_equation}
\nabla \cdot \left( \mathbf{k} \nabla p \right) = \tilde{q}^v_t \mu_1. 
\end{equation}
\noindent The source term $\tilde{q}_t^v$, which depends on the well configuration and the well BHPs, contains a contribution from each well. This term is constructed using Eqs.~\ref{eq:well_rate-eq} and \ref{eq:wellindex-eq}, with $k_{rj}=1$ and $\mu_j=\mu_1$. 

Eq.~\ref{eq:1p_pressure_equation} is relatively inexpensive to solve because it requires only linear solutions in one variable, rather than nonlinear solutions in two variables at many time steps, as is required for Eqs.~\ref{eq:fluid mass balance} and \ref{eq:darcy-eq}. \textcolor{black}{Specifically, for the 2D case considered here, it takes about 1.8~seconds to solve Eq.~\ref{eq:1p_pressure_equation}, compared to about 120~seconds for the solution of Eqs.~\ref{eq:fluid mass balance} and \ref{eq:darcy-eq}. Moreover, as the number of grid blocks increases, the relative cost of solving Eq.~\ref{eq:1p_pressure_equation} versus Eqs.~\ref{eq:fluid mass balance} and \ref{eq:darcy-eq} decreases. After training, a GNSM prediction (including overhead and the single-phase pressure solution) requires about 3.3~seconds. Thus, the computation associated with the solution of Eq.~\ref{eq:1p_pressure_equation} is nontrivial but not overwhelming. In Section~\ref{sec:result_states_rates}, we will quantify the improvement in GNSM accuracy provided by the single-phase flow information.} We note finally that single-phase pressure solutions of the type considered here were also used as a `proxy' for a more complex system in \citep{do2023neural}. In that study, these were applied for well control optimization for two-phase flow in fractured systems.

The edge features, shown in Table~\ref{tab:edge_features}, all involve static quantities. Transmissibility (feature~1) is the finite volume quantity that appears in the discretized flux terms. Specifically, for flow from cell $i$ to cell $j$, transmissibility is given by $T_{i,j}=k_{i,j}A/d_{i,j}$, where $k_{i,j}$ is the harmonic average of cell permeabilities $k_{i}$ and $k_{j}$, $A$ is the area of the shared face between cell $i$ and cell $j$, and $d_{i,j}$ is the distance between the cell centers. The other features appearing in Table~\ref{tab:edge_features} are purely geometric. Dynamic quantities, such as cell-to-cell fluxes and pressure gradients, could be useful in some settings though these are not considered here.

\begin{table}[htb!]
\centering
    \caption{Node features used in the GNSM}
    \begin{tabular}{|c|c|c|}
    \hline
    {No.} & {Feature} & {Quantity} \\ \hline
    {1.} & {${p}^n$} & {current pressure}\\ 
    {2.} & {${S}^n_w$} & {current water saturation}\\ 
    {3.} & {${k}$} & {permeability}\\
    {4.} & {${\phi}$} & {porosity}\\
    {5.} & {${V}$} & {cell bulk volume}\\
    {6.} & {${D}$} & {cell depth}\\
    {7.} & {${W}$} & {well index (for well cells)}\\ 
    {8.} & {${p^w}$} & {wellbore pressure (for well cells)}\\ 
    {9.} & {${e}$} & {encoding of node type}\\ 
    {10.} & {${p}_{1p}$} & {single-phase steady-state pressure}\\
    \hline
    \end{tabular}%
    \label{tab:node_features}%
\end{table}%

\begin{table}[htb!]
\centering
    \caption{Edge features used in the GNSM}
    \begin{tabular}{|c|c|c|}
    \hline
    {No.} & {Feature} & {Quantity} \\ \hline
    {1.} & {${T}$} & {transmissibility}\\ 
    {2.} & {${d}_x$} & {distance in $x$ dimension}\\ 
    {3.} & {${d}_y$} & {distance in $y$ dimension}\\ 
    {4.} & {${d}_z$} & {distance in $z$ dimension}\\ 
    {5.} & {${d}_t$} & {total distance between cells}\\
    \hline
    \end{tabular}%
    \label{tab:edge_features}%
\end{table}%

The hyperparameters considered in the GNSM include those for the MLP, the MPGN, and the encoding-decoding process. Other hyperparameters are associated with training. The full list of hyperparameters is shown in Table~\ref{tab:hyperparameters}. As mentioned previously, we use different GNSMs (SatGNN and PresGNN) for the saturation and pressure state variables. These are tuned separately, and the optimized hyperparameters are, in general, different.

Hyperparameters~1--8 relate to the model architecture. Hyperparameters~1 and 2 define the number of hidden layers and the hidden dimension of the MLPs. These are taken to be the same for all MLPs in both networks. For the MPGN, we consider the number of message passing steps ($N_{msg}$), which defines the set of neighbors from which information is collected to update a particular node, as well as the type of message that is collected (hyperparameters~3 and 4). The latter can involve target node features, neighboring node features, edge features, or differences in node features. Hyperparameter~5 specifies how this information is aggregated (summation, maximization, or averaging). Hyperparameter~6 quantifies the size of the latent space for the encoder and decoder and thus the degree of compression of encoded features compared to the original features. Hyperparameter~7 defines the type of activation layers used, while hyperparameter~8 determines whether and where to use group normalization.

Hyperparameters~9--13 are training related parameters, which we discuss only briefly here (more detail is provided in Section~\ref{sec:training}). Hyperparameter~9 characterizes the Gaussian noise added to the state variables ($p^n$ and $S_w^n$) during training. Hyperparameters~10 and~11 appear as weights in the loss function. Hyperparameter~12 provides the number of training steps used for multistep rollout, while hyperparameter~13 specifies the learning rate. 

\begin{table}[htb!]
\centering
    \caption{Hyperparameters used in the GNSM}
    \begin{tabular}{|c|c|c|}
    \hline
    {No.} & {Hyperparameter} & {Range considered} \\ \hline
    {1.} & {number of hidden layers} & {$2$ or $3$}\\ 
    {2.} & {hidden size} & {$32$, $64$, or $128$}\\ 
    {3.} & {number of message passing layers} & {$3$ -- $15$}\\ 
    {4.} & {type of message passing} & {use difference and edge features or not}\\
    {5.} & {type of aggregation function} & {summation, maximization, or averaging}\\
    {6.} & {latent size} & {$16$ or $32$}\\
    {7.} & {types of activation} & {ReLU, Leaky ReLU, or ELU}\\
    {8.} & {group normalization} & {None, Processor, or MLP}\\
    {9.} & {Gaussian noise std.~dev.}   & {$0.01$, $0.03$, or $0.05$}\\ 
    {10.} & {MAE loss ratio} & {$0.1$ -- $0.9$}\\ 
    {11.} & {well loss ratio} & {$0.1$, $1.0$, or $10.0$}\\ 
    {12.} & {number of multistep training} & {1 -- 6}\\
    {13.} & {learning rate} & {$10^{-2}$, $10^{-3}$, $10^{-4}$, or $10^{-5}$}\\
    \hline
    \end{tabular}%
    \label{tab:hyperparameters}%
\end{table}%

\subsection{Training strategy}
\label{sec:training}

We now describe the procedures used to train the GNSM to predict dynamic state variables. We first introduce the multistage multistep training strategy (MMTS) used in this work. The detailed loss terms will then be presented, and some of the specific treatments will be discussed.

The MMTS applied here is motivated by the multistep training procedure discussed by Wu et al.~\citep{wu2022learning} and the multistage training strategy proposed by Lam et al.~\citep{lam2022graphcast}. Multistage refers to the general approach used for training. With this treatment, different settings and hyperparameters are determined at different training stages. Multistep indicates that the model is trained for successive rollout steps, meaning it will be used (online) to predict over many time steps, with the prediction at time step $n+1$ used as input for step $n+2$, etc. This multistep training process acts to limit error accumulation over multiple steps, which could otherwise lead to unacceptable results. 

Three training stages are used in this study. In the first stage, we use small datasets and evaluate a broad range of hyperparameter combinations, with the goal of identifying promising hyperparameter sets for further evaluation. A total of 
30~samples are used for the unstructured models considered here. Training proceeds for 1000~epochs, and only single-step rollout is considered. The learning rate is a hyperparameter (appearing in Table~\ref{tab:hyperparameters}), which is determined to be $10^{-3}$. This training can be accomplished quickly given the small sample size, even though a large number of epochs is used. 

In the second stage of training, we focus on a few promising sets of hyperparameters. Specifically, six different sets are considered for both the SatGNN and PresGNN. In this stage, training is performed with much larger datasets (600~training simulations are run). Each model is trained for 300~epochs, again for single-step rollout. The learning rate for this stage is determined separately from the first-stage learning rate, though it is again found to be $10^{-3}$.

Multistep training is applied in the third stage. In this stage, we enhance the second-stage models by considering additional time steps, up to a maximum of 6~steps (in total). The learning rate for this stage is determined to be $10^{-5}$, which is much lower than in previous stages. The number of time steps is increased successively (by one) in each training step. We perform 100~epochs for PresGNN and 150~epochs for SatGNN in each of these training steps. This amount of training is sufficient for the training loss to plateau in most cases. 

The error in each step of multistep training contributes to the total loss. This enables SatGNN and PresGNN to avoid exponential growth in error, thus ensuring better long-range predictions. There is a tradeoff, however, between computational cost and prediction accuracy, i.e., increased accuracy can be achieved by considering more steps in multistep training, though this requires more computation and memory. From our numerical experiments, we found that the use of multistep rollout over 6~time steps was computationally tractable while providing reasonable accuracy over the simulation time frame of interest.

Other treatments that proved to be useful during training include the introduction of Gaussian noise and residual prediction. The addition of Gaussian noise was used in~\citep{sanchez2020learning} to improve model robustness against noisy input, and our approach here is similar. Specifically, Gaussian noise with a mean of 0 and prescribed standard deviation (hyperparameter~9 in Table~\ref{tab:hyperparameters}) is added to the normalized training data. This means training is performed using perturbed quantities ${\bf {\tilde P}}^n = {\bf P}^{n} + {\bf \Psi}_p$ and ${\bf \tilde{S}}_w^n = {\bf S}^n_w + {\bf \Psi}_s$, where ${\bf P}^n \in \mathbb{R}^{N_c}$ and ${\bf S}_w^n \in \mathbb{R}^{N_c}$ are the normalized simulation data ($N_c$ is the number of cells), tildes indicate the perturbed data, and ${\bf \Psi}_p \sim \mathcal{N}(0,\,\sigma_p^{2}$) and ${\bf \Psi}_s \sim \mathcal{N}(0,\,\sigma_s^{2})$ denote the Gaussian noise, with $\sigma_p$ and $\sigma_s$ the standard deviations.

With residual prediction, instead of directly predicting the values of the state variables, the GNSM predicts the difference of the state variables between the current step and the next step. Specifically, we predict $\Delta {\bf {\tilde P}}^n = {\bf P}^{n+1}-{\bf {\tilde P}}^{n}$ and $\Delta {\bf \tilde{S}}_w^n = {\bf S}_w^{n+1}-{\bf \tilde{S}}_w^{n}$. Note that we use ${\bf {\tilde P}}^{n}$ and ${\bf {\tilde S}}_w^{n}$ here rather than ${\bf P}^{n}$ and ${\bf S}_w^{n}$.

Both mean squared error (MSE) and mean absolute error (MAE) are used in the loss expression. MSE is more widely applied in deep-learning models, but the MAE loss contribution is also useful as it puts less weight on outliers and has a constant gradient, which can be beneficial for training. We also allow for additional weight to be placed on well blocks. The resulting loss function ($L$) we seek to minimize is

\begin{equation}
    \begin{split}
    L =  &\frac{1}{n_{s} n_t} \left( \sum_{i=1}^{n_{s}}\sum_{k=1}^{n_t} \left\| \Delta \hat{\tilde{\mathbf{X}}}^{k}_{v,i} - \Delta \tilde{\mathbf{X}}^k_{v,i} \right\|_{2}^2 + \alpha \sum_{i=1}^{n_{s}}\sum_{k=1}^{n_t} \left\| \Delta \hat{\tilde{\mathbf{X}}}^{k}_{v,i} - \Delta \tilde{\mathbf{X}}^k_{v,i} \right\|_{1} \right) +  \\
    &\gamma \frac{1}{n_{s} n_t n_w} \left( \sum_{i=1}^{n_{s}}\sum_{k=1}^{n_t}\sum_{j=1}^{n_w} \left\| \delta \hat{\tilde{x}}^{k,j}_{v,i} - \delta \tilde{x}^{k,j}_{v,i} \right\|_2^2 + \beta \sum_{i=1}^{n_{s}}\sum_{k=1}^{n_t}\sum_{j=1}^{n_w} \left\| \delta \hat{\tilde{x}}^{k,j}_{v,i} - \delta \tilde{x}^{k,j}_{v,i} \right\|_1 \right) .
    \end{split}\label{eq:loss_function}
\end{equation}
Here $n_s$ is the number of samples (random well configurations and BHPs) used in training, $n_t$ is the number time steps, $n_w$ is the number of wells, $\alpha$ is the ratio of the MAE loss to MSE loss for matrix nodes (hyperparameter~10), $\beta$ is the analogous quantity for well nodes, and $\gamma$ is the extra weighting for well-cell loss (hyperparameter~11). We use the notation $\Delta {\bf \hat{\tilde X}^\text{$k$}_\text{$v,i$}}$ to indicate the GNSM residual prediction for state variable $v$ ($\Delta {\bf \hat{\tilde X}^\text{$k$}_\text{$p,i$}}$ = $\Delta {\bf \hat{\tilde P}^\text{$k$}_\text{$i$}}$ and $\Delta {\bf \hat{\tilde X}^\text{$k$}_\text{$s,i$}}$ = $\Delta {\bf \hat{\tilde S}^\text{$k$}_{\text{$w$},\text{$i$}}}$) at time step $k$ for sample $i$, and $\Delta {\bf {\tilde X}}^k_{v,i}$ for the corresponding perturbed simulation result. The quantities ${\delta \hat{\tilde x}}^{k,j}_{v,i}$ and ${\delta {\tilde x}}^{k,j}_{v,i}$ are the analogous well-block results, with $j$ indicating the well.

Eq.~\ref{eq:loss_function} is used directly for the first and second stages of MMTS. Multiple steps are considered in the third stage, however, so this stage is treated slightly differently. Specifically, the third-stage loss function is taken to be a weighted sum of the loss for each step. Based on numerical experimentation, we use a weight of 1 for the first step and 0.5 for all subsequent steps. The optimal sets of parameters for PresGNN and SatGNN, corresponding to the minimized loss function after all training stages, are denoted ${\boldsymbol \theta}_v^*$, $v = p, s$.

Well rates are important quantities in many reservoir engineering applications, including optimization. These rates can be computed from the pressure and saturation values in the well blocks through application of Eq.~\ref{eq:well_rate-eq}, with well indices given by Eq.~\ref{eq:wellindex-eq}. We found that this direct approach does not always provide sufficiently accurate well rates, however. In order to improve the accuracy of these estimates, rather than use Eqs.~\ref{eq:well_rate-eq} and \ref{eq:wellindex-eq} directly, we train MLPs to provide well rate predictions from the set of features given in Table~\ref{tab:node_features}. Separate MLPs are trained for water injection rate, oil production rate, and water production rate. The same dataset used to train the GNSM is used for this well-rate training. 

\textcolor{black}{The MLPs used to predict the rates for each well require, as inputs, pressure and saturation in the well region -- specifically, the pressure and saturation in the well block and in all cells that share a face with the well block. There can be a varying number of surrounding cells in unstructured models. These well-region pressure and saturation values are provided by PresGNN and SatGNN. Because the well locations are different for each training sample (or for each candidate solution during optimization), we cannot specify a priori a particular set of grid blocks at which pressure and saturation are required. This means that, in general, pressure and saturation must be determined throughout the model for the well rate computations.}

\begin{table}[htb!]
\centering
    \caption{Optimized GNSM hyperparameters after MMST training}
    \begin{tabular}{|c|c|c|c|}
    \hline
    {No.} & {Hyperparameter} & {PresGNN} & {SatGNN} \\ \hline
    {1.} & {number of hidden layers} & {$2$} & {$2$}\\ 
    {2.} & {hidden size} & {$128$} & {$64$}\\ 
    {3.} & {number of message passing layers} & {$12$} & {$7$}\\ 
    {4.} & {type of message passing} & {self + neighbor info} & {all info}\\
    {5.} & {type of aggregation function} & {summation} & {summation}\\
    {6.} & {latent size} & {$32$} & {$32$}\\
    {7.} & {types of activation} & {ELU} & {ELU}\\
    {8.} & {group normalization} & {None} & {None}\\
    {9.} & {Gaussian noise std.~dev.}   & {$0.03$}   & {$0.01$}\\ 
    {10.} & {MAE loss ratio} & {$0.1$} & {$0.1$}\\ 
    {11.} & {well loss ratio} & {$0.1$} & {$0.1$}\\ 
    {12.} & {number of multistep training} & {6} & {6}\\
    {13.} & {learning rate (stage~1~and~2)} & {$10^{-3}$} & {$10^{-3}$}\\
    {14.} & {learning rate (stage~3)} & {$10^{-5}$} & {$10^{-5}$}\\
    \hline
    \end{tabular}%
    \label{tab:opt_hyperparameters}%
\end{table}%

The final optimized hyperparameters for the PresGNN (${\boldsymbol \theta}_p^*$) and SatGNN (${\boldsymbol \theta}_s^*$) are provided in Table~\ref{tab:opt_hyperparameters}. The time required to train both networks is about 30~hours using a single Nvidia A100 GPU without parallelization, with a batch size of 32. 
This training time is longer (as is often the case for GNN models) than for some other types of surrogate models, such as those based on CNNs. However, as we will see in Section~\ref{sec:result_test}, the GNSM displays a degree of extrapolation capability, meaning it can be applied to new (related) geological models. This will greatly reduce retraining costs when geological uncertainty, represented by considering an ensemble of geological models, is treated. We note finally that training time can be reduced through parallelization, though this was not attempted here. 

\subsection{Online application}
\label{sec:application}

Up to this point, our discussion has focused on the GNSM and the (offline) training process. We now describe the online use of the trained model for new test cases and well placement optimization.

During testing, the initial conditions for the state variables ($p = 200$~bar, $S_w = 0.2$), which are constant over the model, are provided. These correspond to features~1 and 2 in Table~\ref{tab:node_features}. The static node features~3--6, and all edge features listed in Table~\ref{tab:edge_features}, are determined directly from the geomodel. The well configuration and the well controls (BHPs) for each new test sample are randomly generated using the same procedure as was applied to provide the training configurations and BHPs. Given these specifications, $W_i$ is computed for each well from the cell properties (see Eq.~\ref{eq:wellindex-eq}), $p^w$ for each well is applied, and $e$ is formed based on the new well configuration. Thus features~7--9 are determined. As described in Section~\ref{sec:features_and_hyper}, we also need to solve the single-phase steady-state pressure equation (Eq.~\ref{eq:1p_pressure_equation}) for each new configuration. This provides $p_{1p}$ (feature~10).


Given all the required features, we can now predict the system dynamics using rollout. This entails replacing features~1 and~2 at the current step $n$ with the GNSM output state variables (pressure and saturation at step $n+1$). This process is repeated until the end of the simulation time frame is reached.

The GNSM is used similarly for well placement optimization. During optimization, well configurations and BHPs are proposed by the DE algorithm (rather than being generated randomly as is done for testing). The features are then constructed, as described above, for each set of locations and BHPs. The dynamic states are then determined using rollout.

NPV (the objective function to be maximized) must be calculated for each candidate solution. The NPV expression is as follows~\citep{tang2022use}
\begin{equation}\label{eq:npv_def}
    {\text {NPV}} = \sum_{k=1}^{n_t} \frac{\Delta t_k \left[ \sum\limits_{i=1}^{n_{prod}}(p_o q_{o,k}^i - p_{pw} q_{pw,k}^i) - \sum\limits_{i=1}^{n_{inj}} p_{iw} q_{iw,k}^i \right]}{(1+b)^{t_k/365}}.
\end{equation}
Here $n_t$ is the number of time steps (in both the GNSM and simulation), $\Delta t_k$ is the time step size (in days) for step $k$, $t_k$ is the time (in days), and $n_{prod}$ and $n_{inj}$ are the number of production and injection wells ($n_w=n_{prod}+n_{inj}$). The quantity $q_{o,k}^i$ is the oil production rate for well $i$ at time step $k$, $q_{pw,k}^i$ is the water production rate for well $i$ at time step $k$, $q_{iw,k}^i$ is the water injection rate for well $i$ at time step $k$, $p_o$ is the oil price (60~USD/STB in this study), $p_{pw}$ is the cost to treat produced water (3~USD/STB), $p_{iw}$ is the cost of injected water (2~USD/STB), and $b$ is the annual discount rate (here $b=0.1$). All flow rates appearing in Eq.~\ref{eq:npv_def} are in stock tank barrels (STB) per day. 

\textcolor{black}{As explained earlier, well rates are computed using separate MLPs, which take as inputs the GNSM predictions for pressure and saturation at near-well locations. In some optimization problems, the objective function involves global information directly. An example of this in the context of reservoir engineering is the maximization of sweep efficiency; in geological carbon storage, examples include the minimization of mobile CO$_2$ and the minimization of CO$_2$ footprint.}

\section{GNSM performance for test cases}
\label{sec:result_test}
In this section, we first provide more detail on the simulation setup and error quantification. Then, results for performance statistics over the full test set, along with state maps and well rates for particular cases, will be presented. Finally, extrapolation results involving different permeability fields, without any retraining, will be provided. Additional results involving structured-grid cases, with the same BHPs specified for all injectors (and similarly for producers), are presented in Appendix~A.

\subsection{Problem setup and error computation}
\label{sec:result_setup}
\begin{figure*}[!htb]
\centering
\includegraphics[width = 0.5\textwidth]{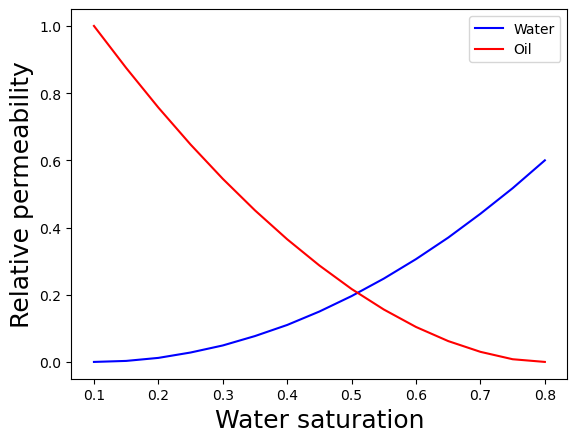}
\caption{Water-oil relative permeability curves used in this study.} \label{fig:rel_perm}
\end{figure*}

The isotropic permeability field and unstructured grid used in all test cases, except where otherwise noted, along with typical well configurations, are shown in Fig.~\ref{fig:setup_example}. This permeability field involves large sand channels (warmer colors) in a background mud (cooler colors). The permeability ranges from 0.39~md to 6552~md. The model covers a physical domain of size of 3000~m $\times$ 3000~m and contains a total of 6045~cells. The reservoir is of thickness 32.8~m. The oil-water relative permeability curves are shown in Fig.~\ref{fig:rel_perm}. The irreducible water saturation is 0.1 and the residual oil saturation is 0.2. Porosity is 0.1 in all cells. The initial pressure in the model is 200~bar and the initial water saturation is 0.2. As noted earlier, in both the training and test cases, the well locations of five injectors and five producers are chosen randomly (with a minimum well-to-well distance of 100~m). Constant-in-time BHPs for each well are randomly sampled from uniform distributions of [210, 310]~bar for injectors and [50, 150]~bar for producers. The simulation time frame is 1500~days. Each simulation time step is 30~days, so there are 50 time steps in total.

We will present test-case results in terms of relative error statistics. Relative error for pressure and saturation for test sample $i$, denoted $e_p^i$ and $e_s^i$, are given by
\begin{equation}\label{eq:rel_err_pressure_sat}
    e_p^i = \frac{1}{n_c n_t} \sum_{j=1}^{n_c} \sum_{t=1}^{n_t} \frac{\left| \hat{P}^t_{i,j} - P^t_{i,j}\right|}{P_{max}-P_{min}}, \ \  e_s^i = \frac{1}{n_c n_t} \sum_{j=1}^{n_c} \sum_{t=1}^{n_t} \frac{\left| \hat{S}^t_{i,j} - S^t_{i,j}\right|}{S^t_{i,j}},
\end{equation}
where $n_c$ is the number of cells in the model, $n_t$ is the number of time steps, $\hat{P}^t_{i,j}$ and $P^t_{i,j}$ are the cell pressures predicted by GNSM and the simulator, respectively, for cell $j$ in test sample $i$ at the time step $t$, and $P_{max}$ and $P_{min}$ are the maximum and minimum pressure in sample $i$ at time step $t$. The quantities $\hat{S}^t_{i,j}$ and $S^t_{i,j}$ are the saturation values from GNSM and the simulator, respectively, for cell $j$ in test sample $i$ at time step $t$. The denominator for saturation error is always nonzero because the initial saturation is prescribed to be 0.2.

Relative error for oil and water production rates, $e_o^i$ and $e_w^i$, are given by
\begin{equation}\label{eq:rel_err_production}
    e_o^i = \frac{1}{N_{prod}} \sum_{j=1}^{N_{prod}} \frac{\int_0^T \left| \hat{q}^o_{i,j}(t) - q^o_{i,j}(t)\right| dt}{\int_0^T \left| q^o_{i,j}(t)\right| dt}, \ \  e_w^i = \frac{1}{N_{prod}} \sum_{j=1}^{N_{prod}} \frac{\int_0^T \left| \hat{q}^w_{i,j}(t) - q^w_{i,j}(t)\right| dt}{\int_0^T \left| q^w_{i,j}(t)\right| dt},
\end{equation}
where $N_{prod}$ is the number of producers, $\hat{q}^o_{i,j}(t)$ and $q^o_{i,j}(t)$ are the oil production rates predicted by GNSM and the simulator in well $j$ for test sample $i$ at time step $t$, and $\hat{q}^w_{i,j}(t)$ and $q^w_{i,j}(t)$ are analogous water production quantities. For injection rate error ($e_{inj}^i$), 
\begin{equation}\label{eq:rel_err_injection}
    e_{inj}^i = \frac{1}{N_{inj}} \sum_{j=1}^{N_{inj}} \frac{\int_0^T \left| \hat{q}^{inj}_{i,j}(t) - q^{inj}_{i,j}(t)\right| dt}{\int_0^T \left| q^{inj}_{i,j}(t)\right| dt},
\end{equation}
where $N_{inj}$ is the number of injectors and $\hat{q}^{inj}_{i,j}(t)$ and $q^{inj}_{i,j}(t)$ are the injection rates from GNSM and the simulator in well $j$ for test sample $i$ at time step $t$.

\subsection{Test set results for states and well rates}
\label{sec:result_states_rates}

\begin{figure*}[!htb]
\centering
\begin{subfigure}{.48\linewidth}\centering
\includegraphics[width=\linewidth]{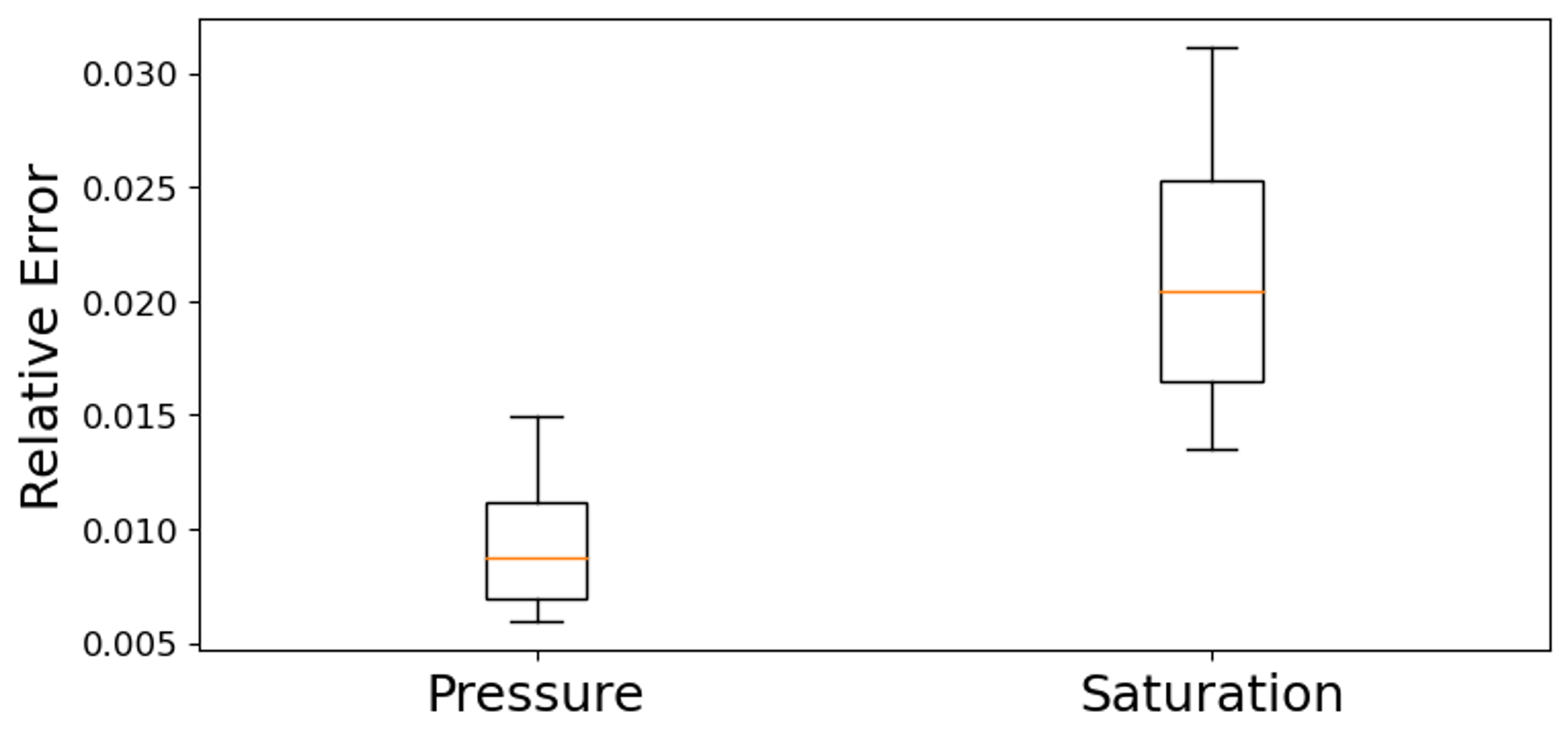}\caption{Box plots of state variables}
\end{subfigure}
\begin{subfigure}{.48\linewidth}\centering
\includegraphics[width=\linewidth]{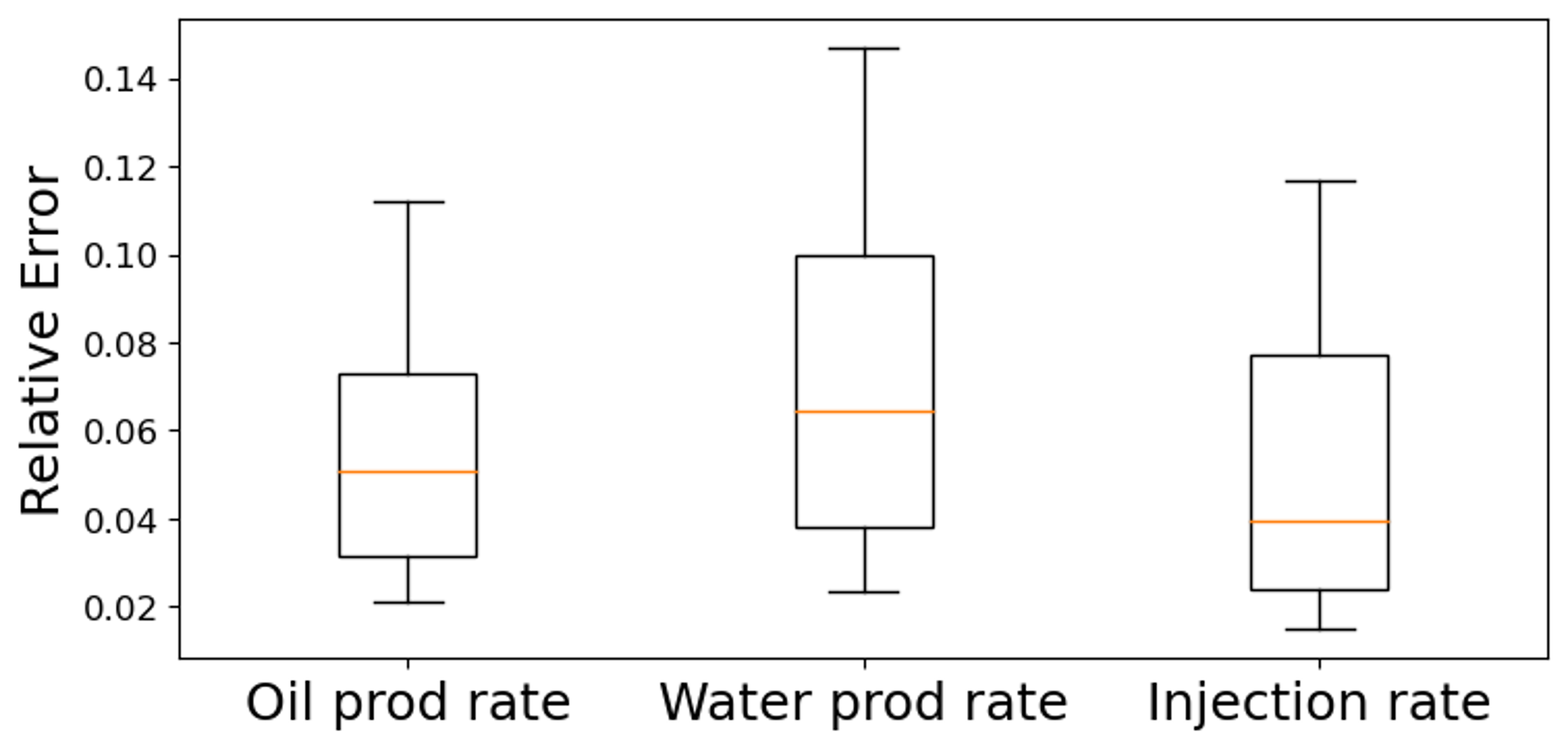}\caption{Box plots of well rates}
\end{subfigure}
\caption{\textcolor{black}{Box plots of relative errors for state variables and well rates over 300 test cases. Boxes display P$_{90}$, P$_{75}$, P$_{50}$, P$_{25}$ and P$_{10}$ (percentile) errors. Error calculations given in Eqs.~\ref{eq:rel_err_pressure_sat}, \ref{eq:rel_err_production} and \ref{eq:rel_err_injection}.}}
\label{fig:state_map_and_well_rates_rel_error}
\end{figure*}

Relative errors over 300 test cases, in terms of box plots, are shown in Fig.~\ref{fig:state_map_and_well_rates_rel_error}. The left plot shows error statistics in the state variables and the right plot shows errors in well rates. In each box, the upper and lower whiskers represent the P$_{90}$ and P$_{10}$ errors, the upper and lower edges of the box denote the P$_{75}$ and P$_{25}$ errors, and the line inside the box shows the P$_{50}$ error. The state variable errors in Fig.~\ref{fig:state_map_and_well_rates_rel_error}(a) are seen to be quite low, with P$_{90}$ errors for pressure and saturation of about 1.5$\%$ and 3$\%$. The oil production, water production, and injection rate errors shown in Fig.~\ref{fig:state_map_and_well_rates_rel_error}(b) are clearly larger than those for the state variables. Here we see P$_{90}$ errors in the 12-14\% range. The median errors, which are about 5$\%$, 7$\%$ and 4$\%$, are moderate. As we will see, the flow rate errors are sufficiently small such that the GNSM can be effectively used for optimization. 

\begin{figure*}[!htb]
\centering
\begin{subfigure}{.32\linewidth}\centering
\includegraphics[width=\linewidth]{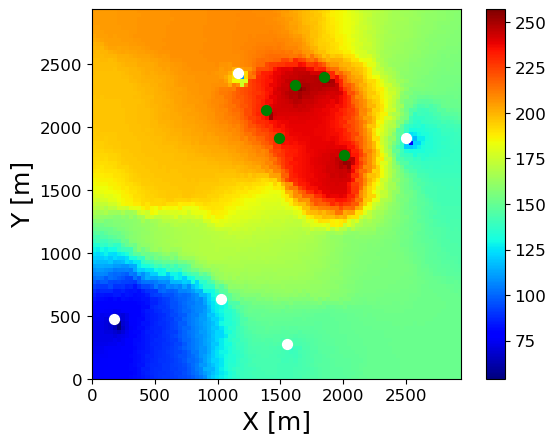}\caption{Simulation pressure}
\end{subfigure}
\begin{subfigure}{.32\linewidth}\centering
\includegraphics[width=\linewidth]{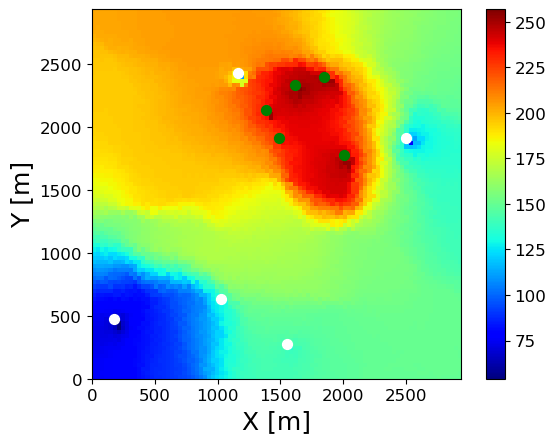}\caption{GNSM pressure}
\end{subfigure}
\begin{subfigure}{.32\linewidth}\centering
\includegraphics[width=\linewidth]{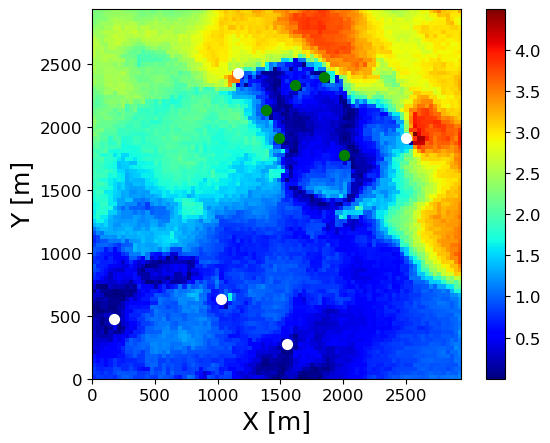}\caption{Pressure difference}
\end{subfigure}
\caption{Pressure maps at 1500~days from simulation (left), GNSM (middle), and their absolute difference (right). Test sample (well configuration and BHPs) here corresponds to the median overall well rate error in the 300~test samples. Green circles indicate injectors and white circles producers (well colors are different than in Fig.~\ref{fig:setup_example} for visual clarity).}
\label{fig:pressure_maps_median_error}
\end{figure*}

\begin{figure*}[!htb]
\centering
\begin{subfigure}{.32\linewidth}\centering
\includegraphics[width=\linewidth]{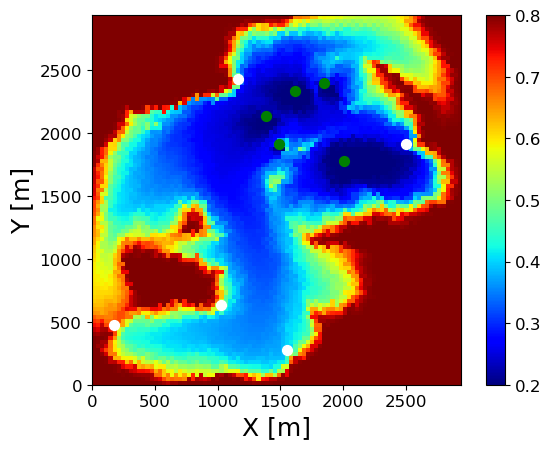}\caption{Simulation saturation}
\end{subfigure}
\begin{subfigure}{.32\linewidth}\centering
\includegraphics[width=\linewidth]{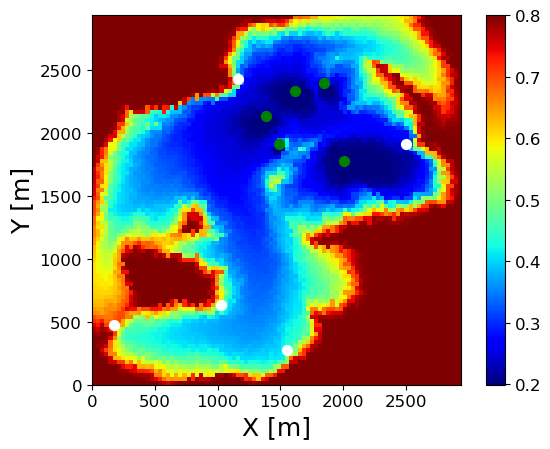}\caption{GNSM saturation}
\end{subfigure}
\begin{subfigure}{.32\linewidth}\centering
\includegraphics[width=\linewidth]{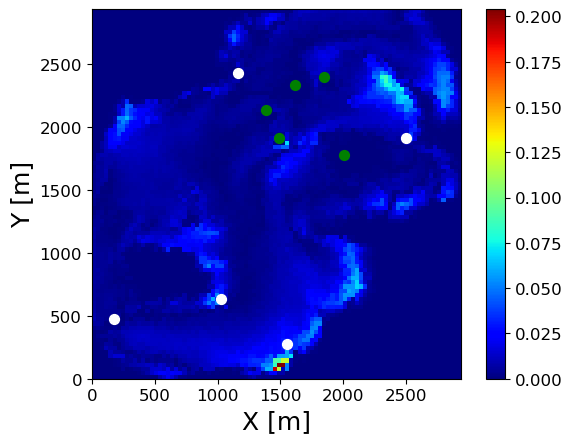}\caption{Saturation difference}
\end{subfigure}
\caption{Saturation maps (blue is water and red is oil) at 1500~days from simulation (left), GNSM (middle), and their absolute difference (right). Test sample (well configuration and BHPs) here corresponds to the median overall well rate error in the 300~test samples. Green circles indicate injectors and white circles producers.}
\label{fig:saturation_maps_median_error}
\end{figure*}

\begin{figure*}[!htb]
\centering
\begin{subfigure}{.32\linewidth}\centering
\includegraphics[width=\linewidth]{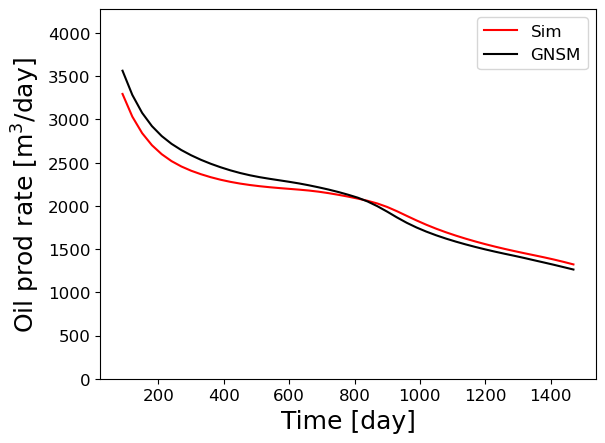}\caption{Field oil rate}
\end{subfigure}
\begin{subfigure}{.32\linewidth}\centering
\includegraphics[width=\linewidth]{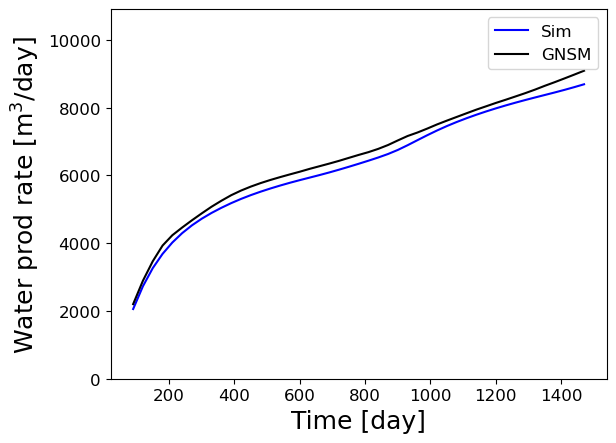}\caption{Field water rate}
\end{subfigure}
\begin{subfigure}{.32\linewidth}\centering
\includegraphics[width=\linewidth]{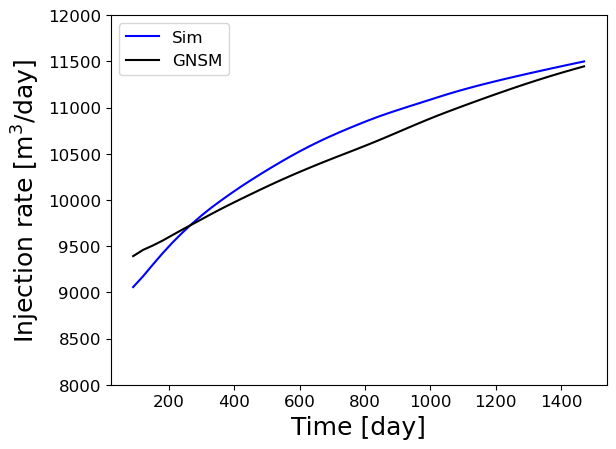}\caption{Field injection rate}
\end{subfigure}
\caption{Field-wide production and injection rates for test sample (well configuration and BHPs) with the median overall well rate error in the 300~test samples.}
\label{fig:P50_rates}
\end{figure*}

\begin{figure*}[!htb]
\centering
\begin{subfigure}{.32\linewidth}\centering
\includegraphics[width=\linewidth]{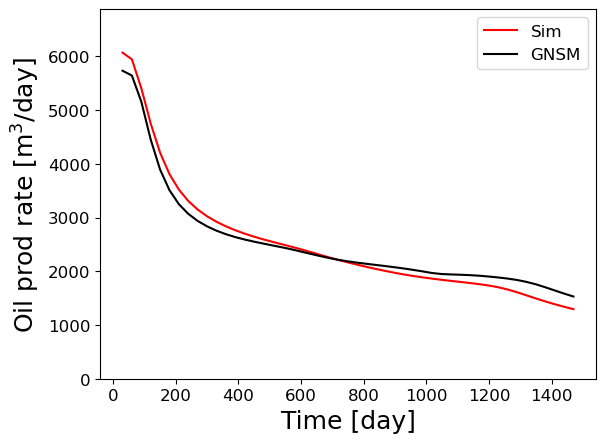}\caption{Field oil rate}
\end{subfigure}
\begin{subfigure}{.32\linewidth}\centering
\includegraphics[width=\linewidth]{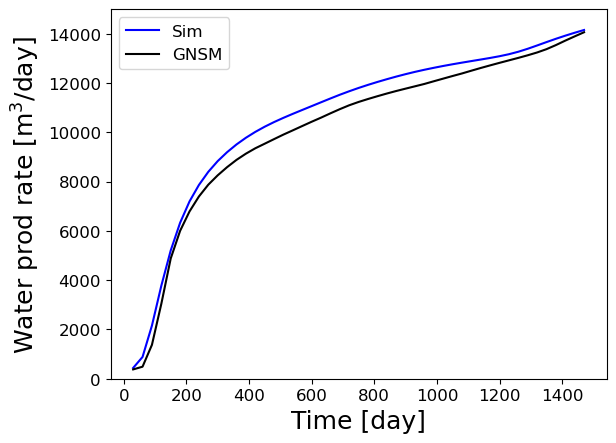}\caption{Field water rate}
\end{subfigure}
\begin{subfigure}{.32\linewidth}\centering
\includegraphics[width=\linewidth]{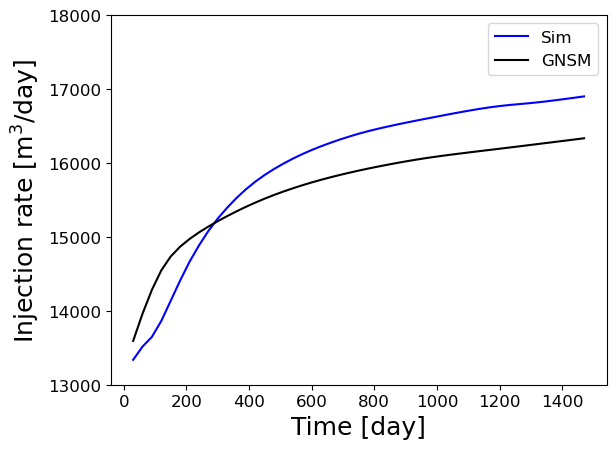}\caption{Field injection rate}
\end{subfigure}
\caption{Field-wide production and injection rates for test sample (well configuration and BHPs) with the P$_{55}$ (i.e., slightly above median) overall well rate error in the 300~test samples.}
\label{fig:P55_rates}
\end{figure*}

Overall well rate error is computed as $(e_o+e_w+e_{inj})/3$. Pressure and saturation maps at 1500~days for the test sample with the median overall well rate error are shown in Figs.~\ref{fig:pressure_maps_median_error} and \ref{fig:saturation_maps_median_error}. Close visual correspondence between the GNSM predictions and reference simulation results is evident in both the pressure and saturation predictions (note the reduced color-bar ranges in the difference plots). In the case of saturation, GNSM errors, though small, appear around the water front.

\begin{figure*}[!htb]
\centering
\begin{subfigure}{.32\linewidth}\centering
\includegraphics[width=\linewidth]{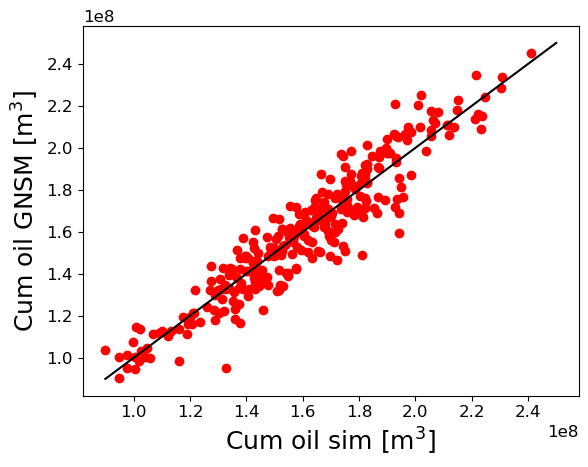}\caption{Cumulative oil}
\end{subfigure}
\begin{subfigure}{.32\linewidth}\centering
\includegraphics[width=\linewidth]{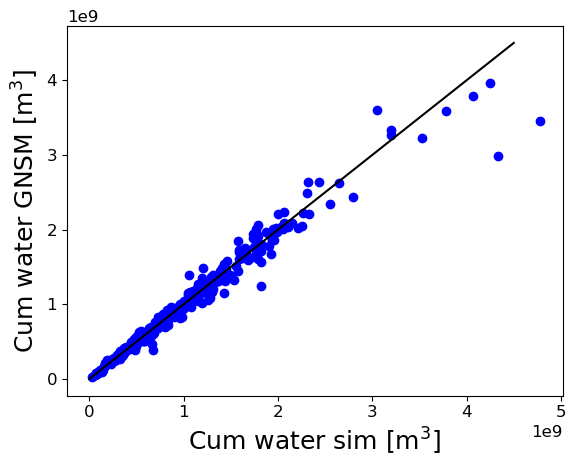}\caption{Cumulative water}
\end{subfigure}
\begin{subfigure}{.32\linewidth}\centering
\includegraphics[width=\linewidth]{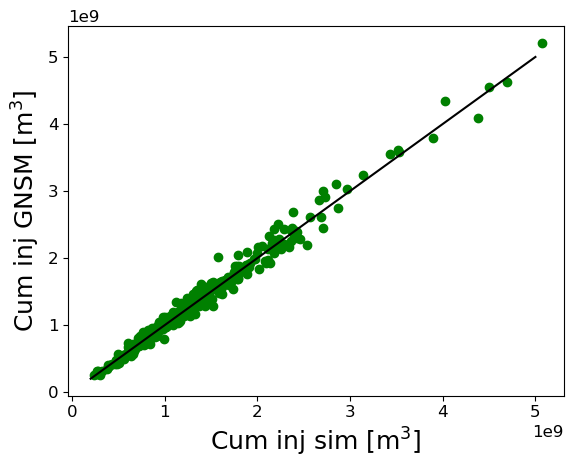}\caption{Cumulative injection}
\end{subfigure}
\caption{Cross-plots of cumulative production and injection quantities. The 45-degree line, corresponding to perfect agreement, is also shown.}
\label{fig:cumulative_rates}
\end{figure*}

Results for field-wide oil rate, water rate, and injection rate are shown in Figs.~\ref{fig:P50_rates} and \ref{fig:P55_rates}. The results in Fig.~\ref{fig:P50_rates} are for the median overall well rate error case (pressure and saturation results for this case are shown in Figs.~\ref{fig:pressure_maps_median_error} and \ref{fig:saturation_maps_median_error}). The field rates provided by GNSM (black curves) for this median error case are seen to be in reasonable agreement with the reference simulation results. Fig.~\ref{fig:P55_rates} displays results for a test sample with overall error (P$_{55}$ case) that is slightly above the median. GNSM predictions for oil and water production rates are accurate, though some error is apparent in the water injection rates. Note the different magnitudes for the various quantities in Figs.~\ref{fig:P50_rates} and \ref{fig:P55_rates}, as well as the different starting values for the field water rate curves. These results demonstrate that the trained GNSM can represent a wide range of flow behaviors.

We next assess GNSM performance in terms of cumulative oil and water produced, and cumulative water injected. These cumulative quantities are time integrations, over the full simulation time frame (1500~days), of the field-wide rates. Cross-plots for these cumulative quantities are presented in Fig.~\ref{fig:cumulative_rates}. The  45-degree lines correspond to perfect agreement. Some scatter is evident in the cumulative oil production plot (Fig.~\ref{fig:cumulative_rates}a), and a few outliers appear in cumulative water production (Fig.~\ref{fig:cumulative_rates}b), though the overall level of agreement in all three quantities is more than adequate. This is an important observation because the NPV computations used in optimization depend strongly on cumulative production and injection. The correspondence is not exact, however, because the NPV computation involves discounting.

\begin{figure*}[!htb]
\centering
\includegraphics[width = 0.75\textwidth]{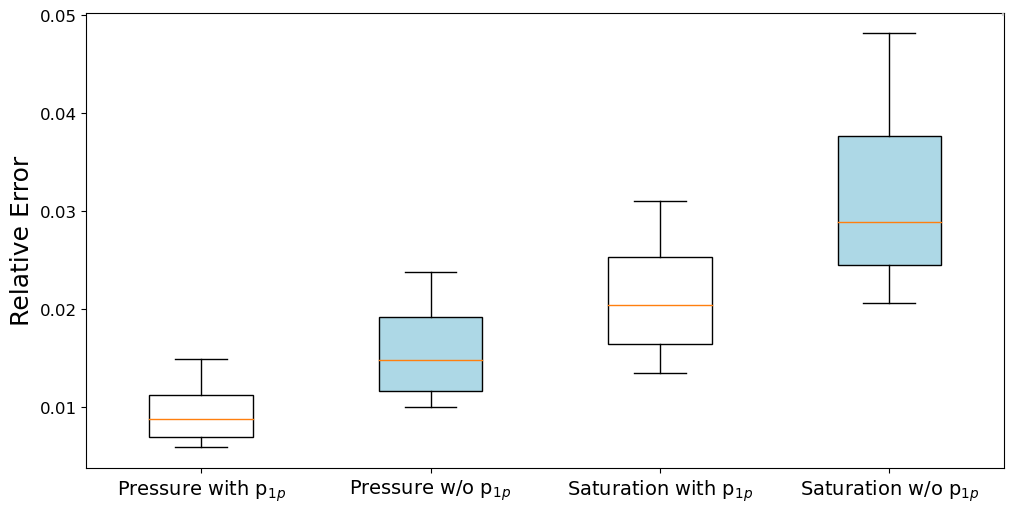}
\caption{\textcolor{black}{Box plots of relative errors for state variables with and without the single-phase steady-state pressure solution as a feature.}} \label{fig:feature_tests}
\end{figure*}

\textcolor{black}{Recall that the use of the single-phase steady-state pressure solution as an input feature adds a computational burden to online GNSM evaluations. We now quantify GNSM accuracy for states and well rates in the absence of this information. The hyperparameters for this assessment are the same as in Table~\ref{tab:opt_hyperparameters}, though the new model was trained for 1.3 times as many epochs, as the loss function required more training to stabilize and flatten. GNSM errors for the 300 test cases, with and without single-phase pressure information, are shown in Fig.~\ref{fig:feature_tests}. GNSM errors without this feature are about 1.5 times larger than those that utilize the feature. Well-rate errors also increase without single-phase pressure information. Specifically, the median error for oil production rate increases from about 5\% to 7.5\%, water production rate from 7\% to 8.5\%, and injection rate from 4\% to 6\%. GNSM accuracy without single-phase pressure information is adequate, overall, and may be acceptable in some applications. This feature is, however, used in all subsequent GNSM results in this paper, since we believe the improved accuracy is worth the extra computation.}


\subsection{GNSM results for new permeability fields}
\label{sec:result_extrap}


\begin{figure*}[!htb]
\centering
\begin{subfigure}{.32\linewidth}\centering
\includegraphics[width=\linewidth]{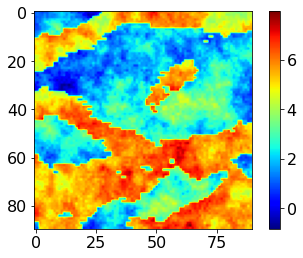}\caption{New permeability field 1}
\end{subfigure}
\begin{subfigure}{.32\linewidth}\centering
\includegraphics[width=\linewidth]{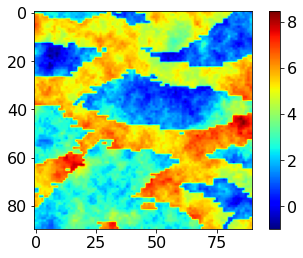}\caption{New permeability field 2}
\end{subfigure}
\begin{subfigure}{.32\linewidth}\centering
\includegraphics[width=\linewidth]{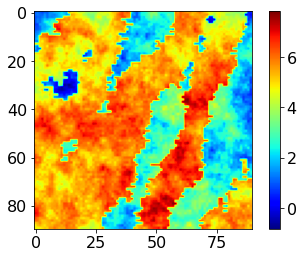}\caption{New permeability field 3}
\end{subfigure}
\caption{\textcolor{black}{New permeability realizations ($\log_e k$ is shown) used for extrapolation tests. These models derive from the same geological training image as the permeability field used for GNSM training (Fig.~\ref{fig:setup_example}), though they involve different geomodel hyperparameters.}}
\label{fig:log_perm_extrapolation}
\end{figure*}

\textcolor{black}{ Finally, we perform `extrapolation' tests of the trained GNSM. For this evaluation, we introduce three new permeability fields. The realizations, shown in Fig.~\ref{fig:log_perm_extrapolation}, were generated with SGeMS from the same geological training image as the permeability field used for the GNSM training simulations (Fig.~\ref{fig:setup_example}). Different geomodel hyperparameters were specified for these realizations. For the permeability realizations in Fig.~\ref{fig:log_perm_extrapolation}(a) and (b), we used a global affinity change in the $y$-direction that was half of that used for the permeability field in Fig.~\ref{fig:setup_example}. This results in channels that are about half as thick. For the geomodel in Fig.~\ref{fig:log_perm_extrapolation}(c), we applied a global rotation of 45 degrees relative to the permeability field in Fig.~\ref{fig:setup_example}. The new realizations are characterized by the same mean and standard deviation of log-permeability, in both the high-permeability channels and low-permeability background regions, as in the original model. }

\textcolor{black} {We perform 100 new test simulations involving randomly generated well configurations and BHPs (over the same ranges as were used previously) for each of the realizations in Fig.~\ref{fig:log_perm_extrapolation}. The other problem specifications are  identical. We reiterate that no additional GNSM training is conducted.}

\begin{figure*}[!htb]
\centering
\includegraphics[width = 0.75\textwidth]{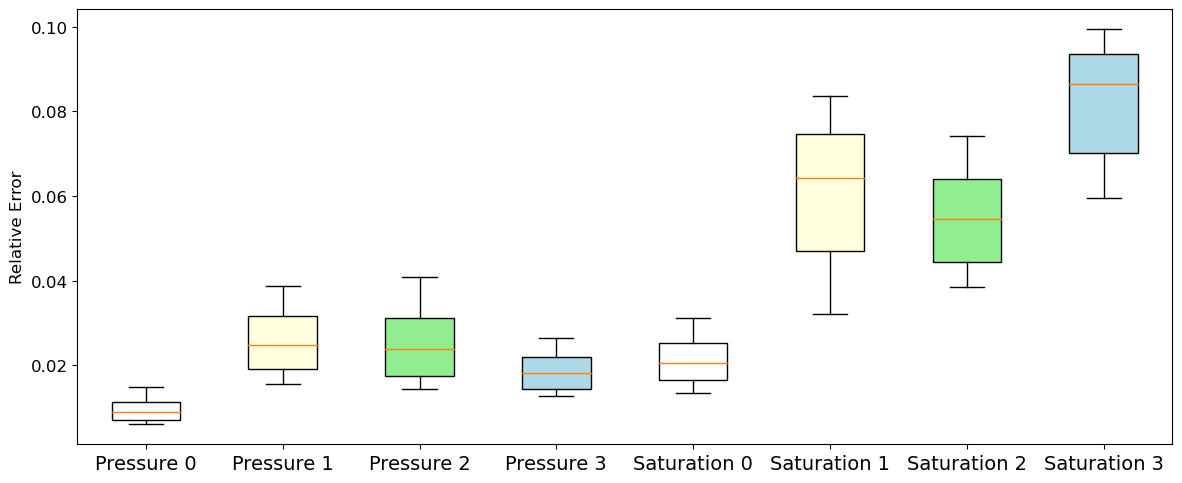}
\caption{\textcolor{black}{ Box plots of relative errors for state variables for different permeability fields. Here 0 refers to the original permeability field (Fig.~\ref{fig:setup_example}, used in training), 1 to the realization in Fig.~\ref{fig:log_perm_extrapolation}(a), 2 to the realization in Fig.~\ref{fig:log_perm_extrapolation}(b), and 3 to the realization in Fig.~\ref{fig:log_perm_extrapolation}(c). GNSM training is performed only for the original model.}} \label{fig:extrapolations_tests}
\end{figure*}

\textcolor{black}{
Box plots of pressure and saturation errors for the new permeability fields, along with errors for the original model (on which training was performed), are presented in Fig.~\ref{fig:extrapolations_tests}. Here 0 refers to the original model (Fig.~\ref{fig:setup_example}), and 1-3 to the three new models in Fig.~\ref{fig:log_perm_extrapolation}. Errors are computed using Eq.~\ref{eq:rel_err_pressure_sat}. It is clear that the pressure and saturation errors are larger for the new permeability fields, as would be expected. Specifically, for pressure, the median errors are 0.010 for the original case, and 0.025, 0.024, and 0.018 (moving from left to right) for the three new models. For saturation, these errors are 0.022 for the original case, and 0.065, 0.055, and 0.086 for the new models. Although the errors clearly increase in these extrapolation cases, the results in Fig.~\ref{fig:extrapolations_tests} demonstrate that the GNSM predictions are still quite reasonable, e.g., the maximum P$_{90}$ relative error for saturation is only about 0.1.}

\begin{figure*}[!htb]
\centering
\begin{subfigure}{.32\linewidth}\centering
\includegraphics[width=\linewidth]{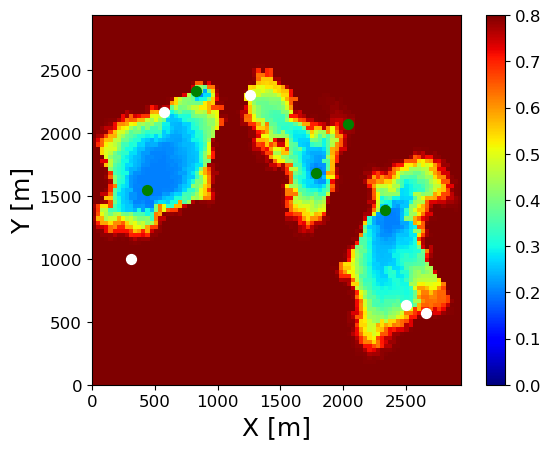}\caption{Simulation saturation}
\end{subfigure}
\begin{subfigure}{.32\linewidth}\centering
\includegraphics[width=\linewidth]{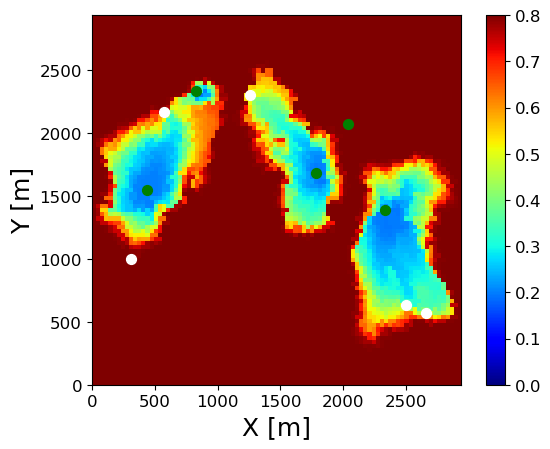}\caption{GNSM saturation}
\end{subfigure}
\begin{subfigure}{.32\linewidth}\centering
\includegraphics[width=\linewidth]{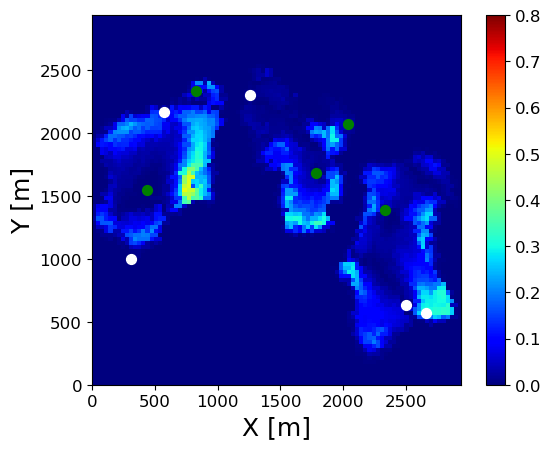}\caption{Saturation difference}
\end{subfigure}
\caption{\textcolor{black}{ Saturation maps at 1500~days from simulation (left), GNSM (middle), and their absolute difference (right) for the new permeability field shown in Fig.~\ref{fig:log_perm_extrapolation}(b) (GNSM trained using permeability realization in Fig.~\ref{fig:setup_example}). Test sample (well configuration and BHPs) here corresponds to the median overall state variable error in the 100~test samples. Green points are injectors and white points are producers.}}
\label{fig:saturation_maps_median_error_extrapolation}
\end{figure*}

\textcolor{black}{
Saturation maps for the test case corresponding to the median overall error in the state variables, for the permeability realization in Fig.~\ref{fig:log_perm_extrapolation}(b), are shown in Fig.~\ref{fig:saturation_maps_median_error_extrapolation}. The simulation and GNSM saturation fields are visually similar, and the errors are mostly near the water front. Overall, the results for the extrapolation test cases in Figs.~\ref{fig:extrapolations_tests} and \ref{fig:saturation_maps_median_error_extrapolation} suggest that the trained GNSM can be applied to different (though related) geomodels. If the goal is to use the GNSM for, e.g., robust optimization over multiple permeability realizations, it will be useful to modify the training procedure such that simulation results for a range of permeability fields are used as training samples. }

\clearpage

\section{Use of GNSM for optimization}
\label{sec:result_opt}
We now apply the GNSM for the optimization of well locations and controls (BHPs). As discussed in Section~\ref{sec:equ_and_opt}, we use a differential evolution (DE) algorithm for the optimization. There are 30 optimization variables, which entail the $x$-$y$ locations for each of the 10 wells, and one BHP value for each well. The population size is also set to 30. The optimization runs are terminated when the relative change in the objective function value is less than 1\% over 15 iterations, or if a total of 50 iterations is reached. In this study, three simulation-based and three GNSM-based optimization runs are conducted. The initial (randomly generated) populations are different in the three runs, but they are the same between the simulation-based and GNSM-based runs. The well-distance constraints and BHP ranges are consistent with those used in the training runs (which are given in Section~\ref{sec:features_and_hyper}). All optimizations are conducted within the Stanford Unified Optimization Framework.

\begin{figure*}[!htb]
\centering
\includegraphics[width = 0.6\textwidth]{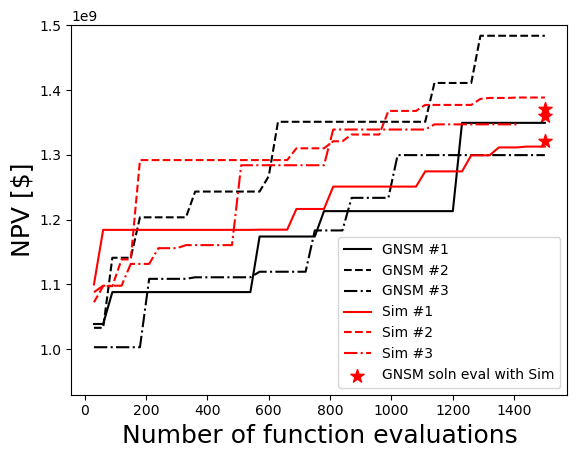}
\caption{Progress of optimizations for simulation-based and GNSM-based runs.} \label{fig:optimization_progress_runs_all_simulation}
\end{figure*}

Fig.~\ref{fig:optimization_progress_runs_all_simulation} displays the progress of the three simulation-based (red curves) and three GNSM-based (black curves) optimization runs. Results are presented in terms of NPV versus the number of function evaluations (flow simulations or GNSM evaluations) performed. \textcolor{black}{Here, the total number of function evaluations is the population size times the number of iterations, both of which can be varied. Consistent with the best-case setup in \citep{zou2022effective}, we take the population size to be equal to the number of unknowns. In general, for the same total number of function evaluations, larger population sizes may not provide better optimization results. In practical applications, it is useful to assess the impact of the various optimizer settings on the target problem. Such an assessment was presented, for DE and other optimizers, in \citep{zou2022effective}. } 

The red stars in Fig.~\ref{fig:optimization_progress_runs_all_simulation} indicate the NPV of the optimal solution (well configuration and BHPs) found by GNSM-based optimization evaluated via simulation. The shifts between the stars and the black curves are indicative of GNSM error at the optimum. Error is significant for GNSM run~2, but it is small for the other two runs. Over the three runs, the best NPV found by simulation-based optimization is $1.39 \times 10^9$~USD, while the best NPV from GNSM-based optimization is $1.37 \times 10^9$~USD. Thus the performance of the two approaches, in terms of objective function value, is close. Note that, because GNSM is much faster to run (timings will be given later), many more GNSM runs than simulation runs can be performed for the same computational budget. Because DE is a stochastic optimization method, it is likely that a higher NPV could be achieved if more runs were performed. This is not done here, however, as our goal is to demonstrate that comparable results can be accomplished with GNSM-based optimization.

\begin{figure*}[!htb]
\centering
\begin{subfigure}{.48\linewidth}\centering
\includegraphics[width=\linewidth]{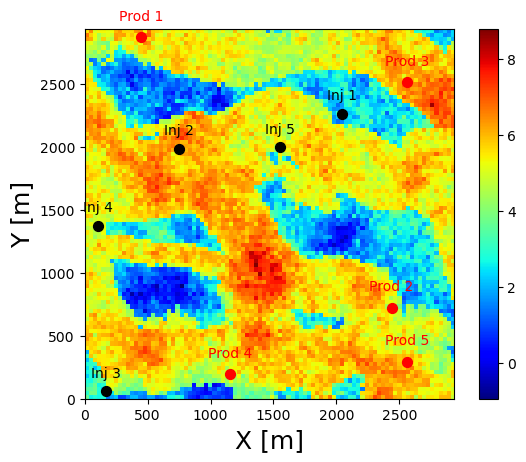}\caption{Optimal locations from simulation}
\end{subfigure}
\begin{subfigure}{.48\linewidth}\centering
\includegraphics[width=\linewidth]{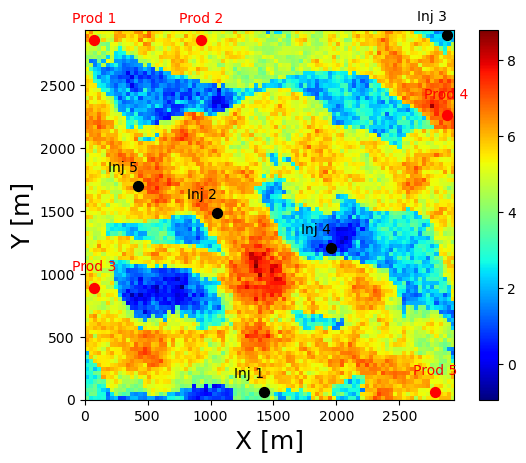}\caption{Optimal locations from GNSM}
\end{subfigure}
\caption{Optimal well configurations from simulation-based and GNSM-based optimization. Black circles indicate injectors and red circles indicate producers. Optimal BHPs given in Table~\ref{tab:bhp_table}.}
\label{fig:optimal_well_configurations}
\end{figure*}

\begin{table}[htb!]
\centering
    \caption{Optimal BHPs from simulation-based and GNSM-based optimization}
    \begin{tabular}{|c|c|c|}
    \hline
    {Well Name} & {Simulation BHP (bar)} & {GNSM BHP (bar)} \\ \hline
    {Inj~1} & {284.0} & {281.0}\\ 
    {Inj~2} & {270.6} & {306.2}\\ 
    {Inj~3} & {264.4} & {248.3}\\ 
    {Inj~4} & {294.8} & {245.1}\\
    {Inj~5}  & {280.8} & {224.0}\\
    {Prd~1} & {61.9} & {61.2}\\ 
    {Prd~2} & {138.5} & {72.6}\\ 
    {Prd~3}  & {50.0} & {150.0} \\
    {Prd~4} & {106.5} & {73.9}\\ 
    {Prd~5} & {128.5} & {89.1}\\ 
    \hline
    \end{tabular}%
    \label{tab:bhp_table}%
\end{table}%

The optimal well locations from both simulation-based and GNSM-based optimization are displayed in Fig.~\ref{fig:optimal_well_configurations}. The optimal BHP value for each well is given in Table~\ref{tab:bhp_table}.  Although the final objective function values for the two solutions are quite similar, the optimal well configurations clearly differ. This is typically observed in this type of optimization and is related to the nonconvexity of the problem, i.e., there are many local optima and the DE algorithm is not expected to find the global minimum. In any event, there are some similarities between the configurations in Fig.~\ref{fig:optimal_well_configurations}(a) and (b). Specifically, both configurations have all five production wells and four of the injection wells in high-permeability channel sand, with the remaining injector in a low-permeability region at the edge of a channel. In addition, both models have three injectors towards the middle of the model, with the other two near the boundaries. The optimal BHPs in Table~\ref{tab:bhp_table} display values within the allowable ranges (210-310~bar for injectors and 50-150~bar for producers) and away from the bounds. This extra degree of freedom enables a wider range of high-quality configurations, as interactions between nearby wells can be `tuned' by varying the BHP values.

\begin{figure*}[!htb]
\centering
\begin{subfigure}{.32\linewidth}\centering
\includegraphics[width=\linewidth]{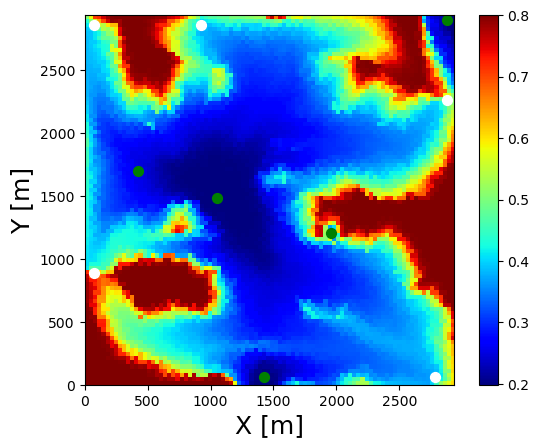}\caption{Simulation saturation}
\end{subfigure}
\begin{subfigure}{.32\linewidth}\centering
\includegraphics[width=\linewidth]{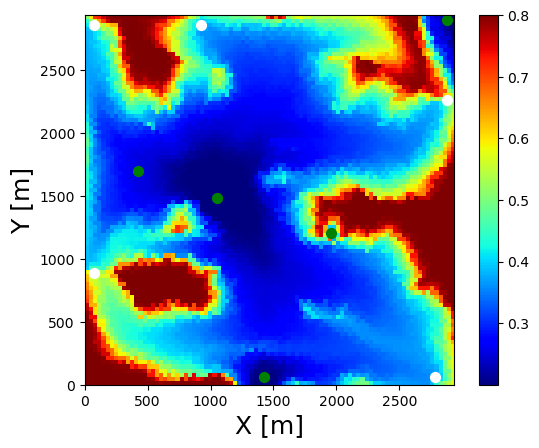}\caption{GNSM saturation}
\end{subfigure}
\begin{subfigure}{.32\linewidth}\centering
\includegraphics[width=\linewidth]{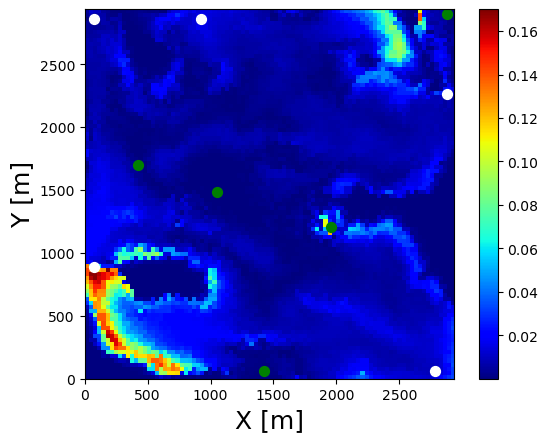}\caption{Saturation differences}
\end{subfigure}
\caption{Saturation maps (blue is water and red is oil) at 1500~days from simulation (left), GNSM (middle), and their absolute difference (right) for the optimal solution found by GNSM-based optimization. Green points are injectors and white points are producers.}
\label{fig:saturation_maps_optimal_gnn}
\end{figure*}

The saturation field corresponding to the optimal solution from GNSM-based optimization (Fig.~\ref{fig:optimal_well_configurations}(b)) is shown in Fig.~\ref{fig:saturation_maps_optimal_gnn}. The left subplot shows simulation results (saturation field at 1500~days) for the solution found by GNSM-based optimization. The GNSM-predicted saturation field appears in the middle subplot, and the difference map is displayed on the right. The GNSM and simulation results are clearly in close agreement, indicating that the optimal GNSM saturation field is accurate. The solutions display a high degree of sweep, with much of the unswept oil located in or near the low-permeability regions evident in Fig.~\ref{fig:optimal_well_configurations}. This is as expected for a solution that maximizes NPV.

\begin{figure*}[!htb]
\centering
\includegraphics[width = 0.7\textwidth]{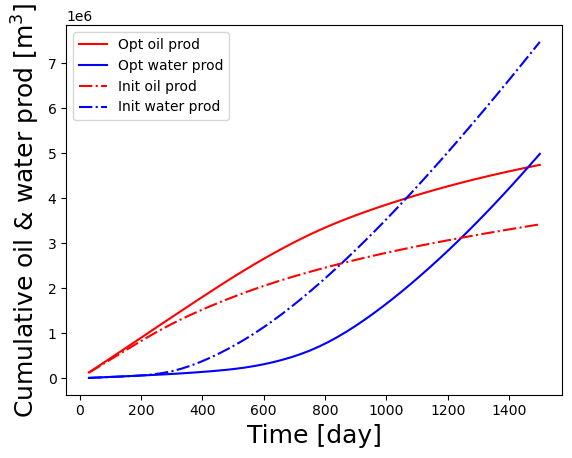}
\caption{Initial and optimal profiles for cumulative oil and water production. Results are for optimal well configuration and BHPs found by GNSM-based optimization, evaluated via simulation.} \label{fig:init_and_opt_oil_water_cumulative}
\end{figure*}

Cumulative oil and water production profiles are shown in Fig.~\ref{fig:init_and_opt_oil_water_cumulative}. Here the cumulative quantity $Q(t)$ is given by $Q(t) = \int_0^t q(\tau) d\tau$, where $q(\tau)$ is the production rate at time $\tau$. The initial solution shown in the figure corresponds to the best solution in the initial DE population. We see that the optimized solution corresponds to more oil production (cumulative oil production of $4.74 \times 10^6$~m$^3$ at the end of the run versus $3.42 \times 10^6$~m$^3$ for the initial solution), along with less water production and later water breakthrough. These behaviors clearly contribute to increased NPV, consistent with the objective of the optimization.

We now discuss the speedup achieved by the GNSM. The ADGPRS simulation runs for this model require about 2~minutes on a single CPU. A GNSM function evaluation takes a total of about 3.3~seconds using a single Nvidia A100 GPU. \textcolor{black}{Of this time, 1.8~seconds are used to solve the single-phase steady-state pressure equation, 0.8~seconds are required for GNSM prediction, and 0.7~seconds are consumed in overhead.} Thus the overall speedup is about a factor of 36. Acceleration of the GNSM evaluations could be achieved by solving the steady-state pressure equation directly (rather than by time stepping) or by parallelizing the GNSM. If we did not use the single-phase steady-state pressure as a feature, GNSM function evaluations would require about 1.5~seconds, and the speedup factor would increase to about 80.  

These timings do not include the offline (preprocessing) associated with the training simulation runs or network training. In the case considered in this paper, we performed 600~training runs (serial time of 20~hours), and GNSM training required about 30~hours. The three optimization runs entailed a combined total of 4500~simulation runs, which corresponds to a total serial time of 150~hours. Overall speedup will be enhanced if more GNSM-based optimizations are performed. More dramatic efficiency gains may be achieved in robust optimization settings, where each candidate solution proposed by the optimizer must be evaluated over a set of geological realizations. In this case GNSM training costs will be quite small relative to the cost of simulation-based optimization.


\section{Concluding remarks}
\label{sec:summary}
In this study, we developed a graph network surrogate model (GNSM) to predict flow responses for cases with shifting well locations and different (constant-in-time) well BHPs in unstructured-grid systems. The model employs an encoding-processing-decoding architecture, where the reservoir model is transformed into a computational graph in which the state variables are updated in time. We use different graph networks, referred to as PresGNN and SatGNN, for the pressure and saturation variables. Improved accuracy in states and well rates was achieved by incorporating the single-phase steady-state pressure solution as a feature and by introducing loss terms for well cells. To improve the predictive capability over multiple time steps, a multistage multistep training strategy (MMTS) was adopted. This approach acts to limit error accumulation over multiple time steps. The two networks are trained independently, with different hyperparameters determined for each. 

The trained GNSM was tested for cases involving five injection wells and five production wells placed randomly in a 2D unstructured channelized model. The well BHPs were also assigned randomly (within specified ranges). For a set of 300 test cases, the GNSM achieved median errors for the pressure and saturation states of approximately 1$\%$ and 2$\%$, respectively. Median errors were slightly higher for well rates -- 5$\%$ for oil production, 7$\%$ for water production, and 4$\%$ for water injection. 

\textcolor{black}{An extrapolation assessment, in which the trained model was evaluated for test cases involving three new permeability realizations (which clearly differ from the original model), was also presented. GNSM predictions for the new permeability fields were less accurate than those for the original model (used for training), though the results were still of reasonable accuracy. This demonstrates a degree of extrapolation capability in the trained GNSM, which could be very useful for some applications.} Finally, the surrogate model was used for optimization, where the goal was to determine the optimal locations and BHP settings such that the NPV of a waterflood operation was maximized. Results using GNSM-based optimization were close to those from simulation-based optimizations, indicating the applicability of the method for this important application. A runtime speedup of about a factor of 36 was achieved for these optimizations.

There are many directions that should be explored in future work. The models considered in this study are relatively small and 2D. The methodology should be tested and extended as necessary to handle larger and more realistic 3D cases. Additional control variables, e.g., well BHPs or rates at a set of control steps, should also be incorporated. The use of the GNSM for robust optimization, where expected NPV is maximized over multiple geological realizations, should be tested. This will require enhancement of the GNSM extrapolation capability. More advanced graph neural network architectures, including graph transformers and graph U-Net, should be considered. These could optimize neuron usage, potentially leading to higher accuracy and/or computational speedup. Finally, the model should be extended to other subsurface flow settings, including geological carbon storage. Here the goal could be to maximize pore space utilization or to minimize risks associated with the storage operation. \textcolor{black}{In such settings, global state information, which is provided by the GNSM, is required to evaluate the objective function.}

\section*{CRediT authorship contribution statement}
\textbf{Haoyu Tang}: Conceptualization, Methodology, Coding, Generation of results, Results interpretation, Visualization, Writing – original draft. \textbf{Louis J. Durlofsky}: Conceptualization, Supervision, Project administration, Funding acquisition, Results interpretation, Resources, Writing – review \& editing.

\section*{Declaration of competing interest}
The authors declare that they have no known competing financial interests or personal relationships that could have appeared to influence the work reported in this paper.

\section*{Declaration of generative AI use in the writing process}
In writing the initial draft of this paper, ChatGPT was used for grammar-related issues and to improve wording. The authors reviewed and edited the paper and take full responsibility for its content.

\section*{Acknowledgements}
We are grateful to the Stanford Smart Fields Consortium for funding and to the SDSS Center for Computation for computational resources. We thank Oleg Volkov and Amy Zou for their assistance with the Stanford Unified Optimization Framework, and Su Jiang for providing geological models.

\bibliography{ref}

\begin{thebibliography}{10}
\expandafter\ifx\csname url\endcsname\relax
  \def\url#1{\texttt{#1}}\fi
\expandafter\ifx\csname urlprefix\endcsname\relax\def\urlprefix{URL }\fi
\expandafter\ifx\csname href\endcsname\relax
  \def\href#1#2{#2} \def\path#1{#1}\fi

\bibitem{chai2021integrated}
Z.~Chai, A.~Nwachukwu, Y.~Zagayevskiy, S.~Amini, S.~Madasu, An integrated closed-loop solution to assisted history matching and field optimization with machine learning techniques, Journal of Petroleum Science and Engineering 198 (2021) 108--204.

\bibitem{kim2023convolutional}
Y.~D. Kim, L.~J. Durlofsky, Convolutional--recurrent neural network proxy for robust optimization and closed-loop reservoir management, Computational Geosciences 27 (2023) 179--202.

\bibitem{nasir2023deep}
Y.~Nasir, L.~J. Durlofsky, Deep reinforcement learning for optimizing well settings in subsurface systems with uncertain geology, Journal of Computational Physics 477 (2023) 111945.

\bibitem{xu2022uncertainty}
R.~Xu, D.~Zhang, N.~Wang, Uncertainty quantification and inverse modeling for subsurface flow in 3{D} heterogeneous formations using a theory-guided convolutional encoder-decoder network, Journal of Hydrology 613 (2022) 128321.

\bibitem{wang2022surrogate}
N.~Wang, H.~Chang, D.~Zhang, Surrogate and inverse modeling for two-phase flow in porous media via theory-guided convolutional neural network, Journal of Computational Physics 466 (2022) 111419.

\bibitem{tang2021deep}
M.~Tang, Y.~Liu, L.~J. Durlofsky, Deep-learning-based surrogate flow modeling and geological parameterization for data assimilation in 3{D} subsurface flow, Computer Methods in Applied Mechanics and Engineering 376 (2021) 113636.

\bibitem{geneva2020modeling}
N.~Geneva, N.~Zabaras, Modeling the dynamics of {PDE} systems with physics-constrained deep auto-regressive networks, Journal of Computational Physics 403 (2020) 109056.

\bibitem{razak2022conditioning}
S.~M. Razak, B.~Jafarpour, Conditioning generative adversarial networks on nonlinear data for subsurface flow model calibration and uncertainty quantification, Computational Geosciences 26~(1) (2022) 29--52.

\bibitem{tang2022deep}
M.~Tang, X.~Ju, L.~J. Durlofsky, Deep-learning-based coupled flow-geomechanics surrogate model for {CO$_2$} sequestration, International Journal of Greenhouse Gas Control 118 (2022) 103692.

\bibitem{wen2022accelerating}
G.~Wen, Z.~Li, Q.~Long, K.~Azizzadenesheli, A.~Anandkumar, S.~M. Benson, Accelerating carbon capture and storage modeling using {F}ourier neural operators, arXiv preprint arXiv:2210.17051, 2022.

\bibitem{grady2022towards}
T.~J. Grady~II, R.~Khan, M.~Louboutin, Z.~Yin, P.~A. Witte, R.~Chandra, R.~J. Hewett, F.~J. Herrmann, Towards large-scale learned solvers for parametric {PDE}s with model-parallel {F}ourier neural operators, arXiv preprint arXiv:2204.01205, 2022.

\bibitem{kim2020robust}
J.~Kim, H.~Yang, J.~Choe, Robust optimization of the locations and types of multiple wells using {CNN} based proxy models, Journal of Petroleum Science and Engineering 193 (2020) 107424.

\bibitem{tang2022use}
H.~Tang, L.~J. Durlofsky, Use of low-fidelity models with machine-learning error correction for well placement optimization, Computational Geosciences 26~(5) (2022) 1189--1206.

\bibitem{nwachukwu2018fast}
A.~Nwachukwu, H.~Jeong, M.~Pyrcz, L.~W. Lake, Fast evaluation of well placements in heterogeneous reservoir models using machine learning, Journal of Petroleum Science and Engineering 163 (2018) 463--475.

\bibitem{mousavi2020optimal}
S.~M. Mousavi, H.~Jabbari, M.~Darab, M.~Nourani, S.~Sadeghnejad, Optimal well placement using machine learning methods: Multiple reservoir scenarios, in: SPE Norway Subsurface Conference, OnePetro, 2020.

\bibitem{redouane2018automated}
K.~Redouane, N.~Zeraibi, M.~Nait~Amar, Automated optimization of well placement via adaptive space-filling surrogate modelling and evolutionary algorithm, in: Abu Dhabi International Petroleum Exhibition \& Conference, OnePetro, 2018.

\bibitem{bruyelle2019well}
J.~Bruyelle, D.~Gu{\'e}rillot, Well placement optimization with an artificial intelligence method applied to {B}rugge field, in: SPE Gas \& Oil Technology Showcase and Conference, OnePetro, 2019.

\bibitem{wang2022efficient}
N.~Wang, H.~Chang, D.~Zhang, L.~Xue, Y.~Chen, Efficient well placement optimization based on theory-guided convolutional neural network, Journal of Petroleum Science and Engineering 208 (2022) 109545.

\bibitem{sanchez2020learning}
A.~Sanchez-Gonzalez, J.~Godwin, T.~Pfaff, R.~Ying, J.~Leskovec, P.~Battaglia, Learning to simulate complex physics with graph networks, in: International Conference on Machine Learning, PMLR, 2020, pp. 8459--8468.

\bibitem{pfaff2020learning}
T.~Pfaff, M.~Fortunato, A.~Sanchez-Gonzalez, P.~W. Battaglia, Learning mesh-based simulation with graph networks, arXiv preprint arXiv:2010.03409, 2020.

\bibitem{wulearning}
T.~Wu, T.~Maruyama, Q.~Zhao, G.~Wetzstein, J.~Leskovec, Learning controllable adaptive simulation for multi-resolution physics, in: The Eleventh International Conference on Learning Representations, 2023.

\bibitem{zhao2022learning}
Q.~Zhao, D.~B. Lindell, G.~Wetzstein, Learning to solve {PDE}-constrained inverse problems with graph networks, arXiv preprint arXiv:2206.00711, 2022.

\bibitem{li2022graph}
Z.~Li, A.~B. Farimani, Graph neural network-accelerated {L}agrangian fluid simulation, Computers \& Graphics 103 (2022) 201--211.

\bibitem{belbute2020combining}
F.~D.~A. Belbute-Peres, T.~Economon, Z.~Kolter, Combining differentiable {PDE} solvers and graph neural networks for fluid flow prediction, in: International Conference on Machine Learning, PMLR, 2020, pp. 2402--2411.

\bibitem{lienen2022learning}
M.~Lienen, S.~G{\"u}nnemann, Learning the dynamics of physical systems from sparse observations with finite element networks, arXiv preprint arXiv:2203.08852, 2022.

\bibitem{lam2022graphcast}
R.~Lam, A.~Sanchez-Gonzalez, M.~Willson, P.~Wirnsberger, M.~Fortunato, A.~Pritzel, S.~Ravuri, T.~Ewalds, F.~Alet, Z.~Eaton-Rosen, et~al., Graphcast: Learning skillful medium-range global weather forecasting, arXiv preprint arXiv:2212.12794, 2022.

\bibitem{wu2022learning}
T.~Wu, Q.~Wang, Y.~Zhang, R.~Ying, K.~Cao, R.~Sosic, R.~Jalali, H.~Hamam, M.~Maucec, J.~Leskovec, Learning large-scale subsurface simulations with a hybrid graph network simulator, in: Proceedings of the 28th ACM SIGKDD Conference on Knowledge Discovery and Data Mining, 2022, pp. 4184--4194.

\bibitem{peaceman1983interpretation}
D.~W. Peaceman, Interpretation of well-block pressures in numerical reservoir simulation with nonsquare grid blocks and anisotropic permeability, Society of Petroleum Engineers Journal 23~(03) (1983) 531--543.

\bibitem{lie2019introduction}
K.-A. Lie, An Introduction to Reservoir Simulation Using MATLAB/GNU Octave: User Guide for the MATLAB Reservoir Simulation Toolbox (MRST), Cambridge University Press, 2019.

\bibitem{zhou2012parallel}
Y.~Zhou, Parallel general-purpose reservoir simulation with coupled reservoir models and multisegment wells, Ph.D. thesis, Stanford University (2012).

\bibitem{price2006differential}
K.~Price, R.~M. Storn, J.~A. Lampinen, Differential Evolution: A Practical Approach to Global Optimization, Springer Science \& Business Media, 2006.

\bibitem{zou2022effective}
A.~Zou, T.~Ye, O.~Volkov, L.~J. Durlofsky, Effective treatment of geometric constraints in derivative-free well placement optimization, Journal of Petroleum Science and Engineering 215 (2022) 110635.

\bibitem{do2023neural}
Y.~Do~Kim, L.~J. Durlofsky, Neural network surrogate for flow prediction and robust optimization in fractured reservoir systems, Fuel 351 (2023) 128756.

\end{thebibliography}

\section*{Appendix~A.~GNSM performance for structured-grid cases}

In this appendix, we apply our GNSM to a structured-model case. The formulation is further simplified by specifying that all injection wells operate under constant-in-time BHPs of 300~bar, and all production wells operate at 100~bar. Results for this simpler setup are useful as they allow us to assess whether the GNSM degrades for the more challenging unstructured cases, with different BHPs, considered in the main text. The structured model uses the same permeability field, shown in Fig.~\ref{fig:setup_example}, mapped to a $90 \times 90$ Cartesian grid. The other simulation settings are the same as in Section~\ref{sec:result_test}. The GNSM hyperparameters must be determined separately for this case. These are similar to those in Table~\ref{tab:opt_hyperparameters}, except now the hidden size for SatGNN is 128, the number of message passing layers for SatGNN is 5, the activation for SatGNN is Leaky ReLU, and the aggregation function for PresGNN is maximization.

\begin{figure*}[!htb]
\centering
\begin{subfigure}{.48\linewidth}\centering
\includegraphics[width=\linewidth]{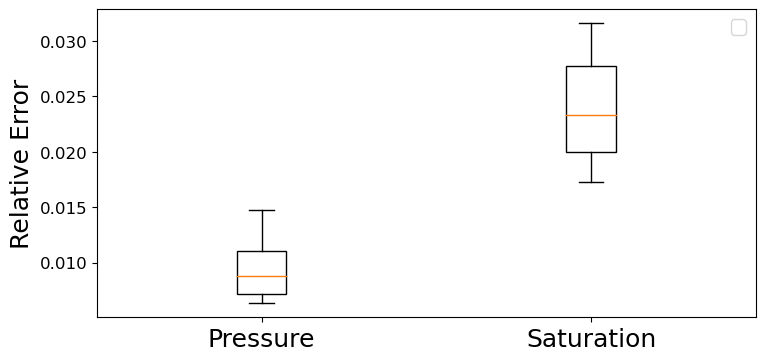}\caption{Box plots of state variables}
\end{subfigure}
\begin{subfigure}{.48\linewidth}\centering
\includegraphics[width=\linewidth]{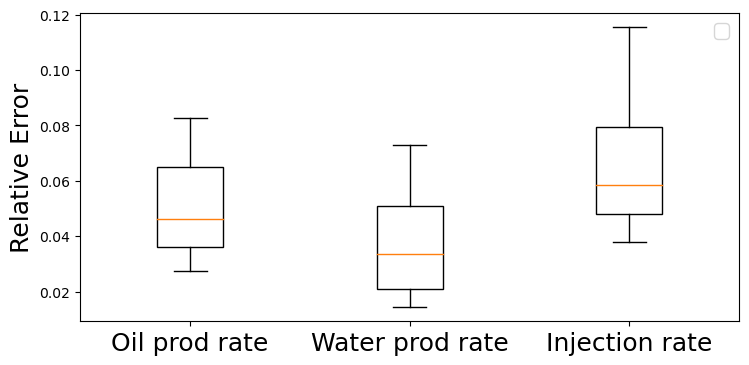}\caption{Box plots of well rates}
\end{subfigure}
\caption{Box plots of relative errors for state variables and well rates over 300 structured-grid test cases. Error calculations given in Eqs.~\ref{eq:rel_err_pressure_sat}, \ref{eq:rel_err_production} and \ref{eq:rel_err_injection}.}
\label{fig:state_map_and_well_rates_rel_error_structured}
\end{figure*}

\begin{figure*}[!htb]
\centering
\begin{subfigure}{.32\linewidth}\centering
\includegraphics[width=\linewidth]{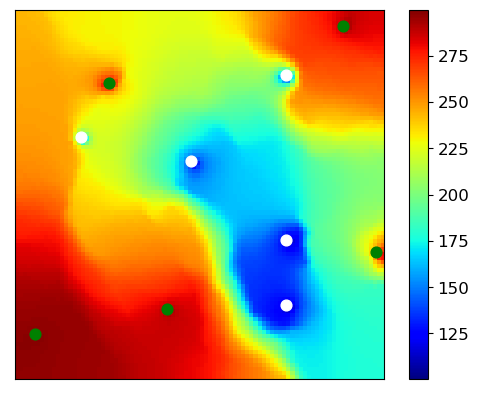}\caption{Simulation pressure}
\end{subfigure}
\begin{subfigure}{.32\linewidth}\centering
\includegraphics[width=\linewidth]{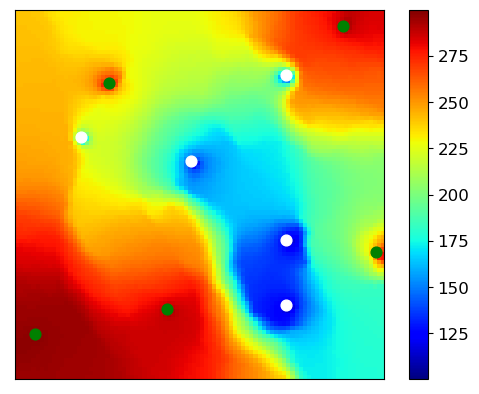}\caption{GNSM pressure}
\end{subfigure}
\begin{subfigure}{.32\linewidth}\centering
\includegraphics[width=\linewidth]{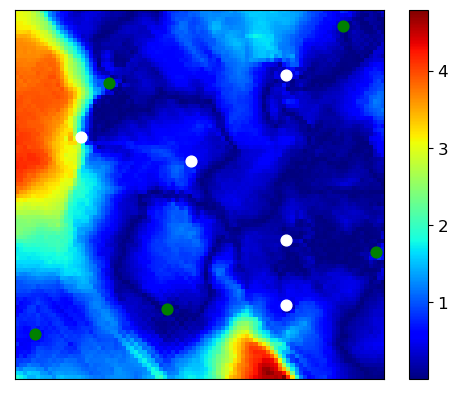}\caption{Pressure difference}
\end{subfigure}
\caption{Pressure maps at 1500~days from simulation (left), GNSM (middle), and their absolute difference (right). Test sample (well configuration) here corresponds to the median overall well rate error in the 300 structured-grid test cases. Green points are injectors and white points are producers.}
\label{fig:pressure_maps_median_error_structured}
\end{figure*}

\begin{figure*}[!htb]
\centering
\begin{subfigure}{.32\linewidth}\centering
\includegraphics[width=\linewidth]{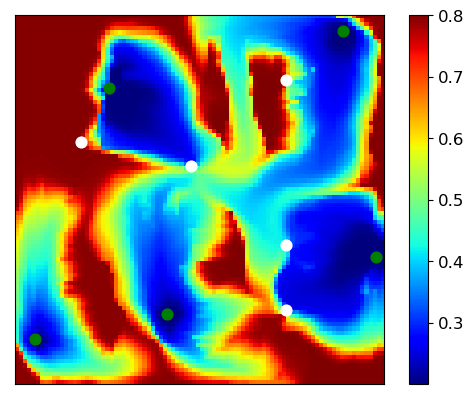}\caption{Simulation saturation}
\end{subfigure}
\begin{subfigure}{.32\linewidth}\centering
\includegraphics[width=\linewidth]{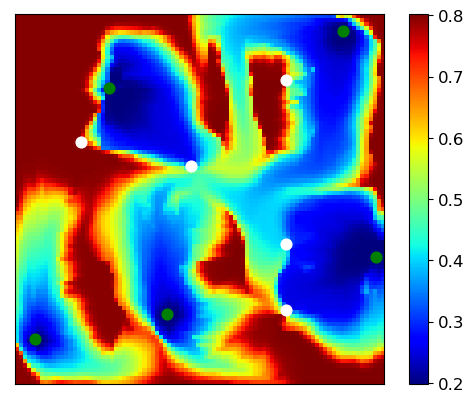}\caption{GNSM saturation}
\end{subfigure}
\begin{subfigure}{.32\linewidth}\centering
\includegraphics[width=\linewidth]{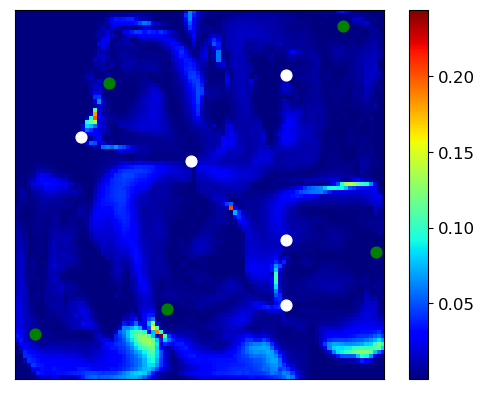}\caption{Saturation difference}
\end{subfigure}
\caption{Saturation maps (blue is water and red is oil) at 1500~days from simulation (left), GNSM (middle), and their absolute difference (right). Test sample (well configuration) here corresponds to the median overall well rate error in the 300 structured-grid test cases. Green points are injectors and white points are producers.}
\label{fig:saturation_maps_median_error_structured}
\end{figure*}

Box plots of the relative errors over 300 test cases (each of which corresponds to a new random well configuration with five injectors and five producers) for the state variables and well rates are presented in Fig.~\ref{fig:state_map_and_well_rates_rel_error_structured}. These errors are comparable to those for the unstructured case, shown in Fig.~\ref{fig:state_map_and_well_rates_rel_error}. Specifically, median errors for pressure and saturation here are 0.83$\%$ and 2.3$\%$, while for the unstructured case (Fig.~\ref{fig:state_map_and_well_rates_rel_error}(a)) these errors are 1.0$\%$ and 2.2$\%$. Median well rate errors in Fig.~\ref{fig:state_map_and_well_rates_rel_error_structured}(b) are 4.7\%, 3.3\% and 5.9\%. For the unstructured case (Fig.~\ref{fig:state_map_and_well_rates_rel_error}(b)), these errors are 5.1\%, 6.5\% and 4.0\%. 

\begin{figure*}[!htb]
\centering
\begin{subfigure}{.32\linewidth}\centering
\includegraphics[width=\linewidth]{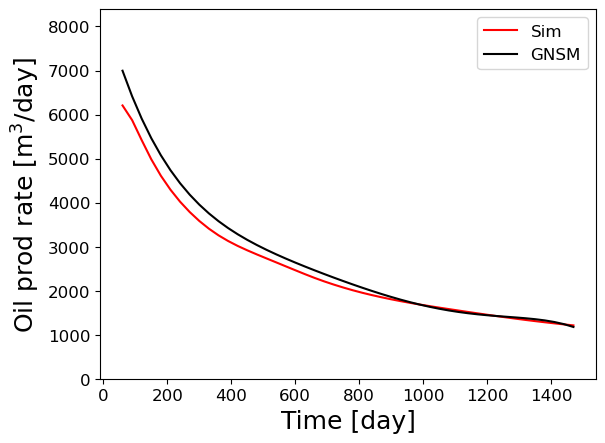}\caption{Field oil rate}
\end{subfigure}
\begin{subfigure}{.32\linewidth}\centering
\includegraphics[width=\linewidth]{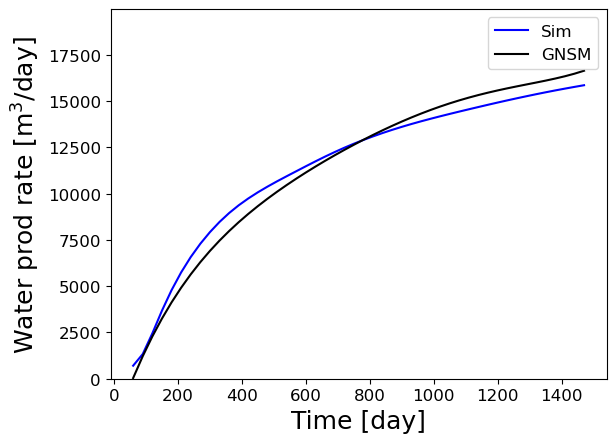}\caption{Field water rate}
\end{subfigure}
\begin{subfigure}{.32\linewidth}\centering
\includegraphics[width=\linewidth]{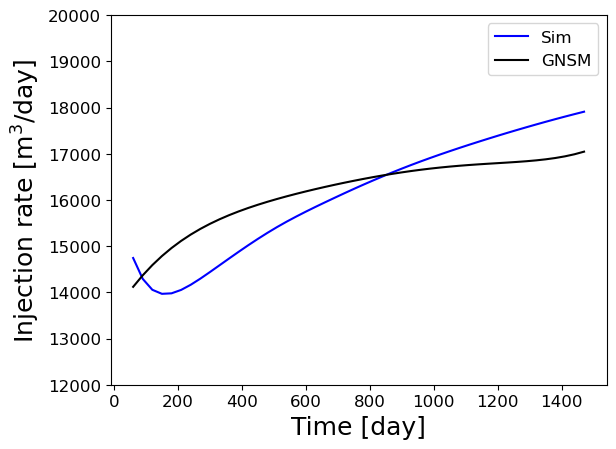}\caption{Field injection rate}
\end{subfigure}
\caption{Field oil rate (left), water rate (middle), and injection rate (right) corresponding to the median overall well rate error over 300 structured-grid test cases.}
\label{fig:P50_rates_structured}
\end{figure*}

Pressure and saturation maps at 1500~days for the structured-case test sample with the median overall well rate error are shown in Figs.~\ref{fig:pressure_maps_median_error_structured} and \ref{fig:saturation_maps_median_error_structured}. The GNSM results are in close visual agreement with the simulation results, and the difference maps are in the same ranges as those in Figs.~\ref{fig:pressure_maps_median_error} and \ref{fig:saturation_maps_median_error}. GNSM predictions for oil and water production and water injection rates are shown in Fig.~\ref{fig:P50_rates_structured}. The level of agreement in the oil and water production results is similar to that in Fig.~\ref{fig:P50_rates}, though water injection here shows larger error.

In total, we observe comparable overall results between the unstructured case (with different well BHPs) and the structured case (with the same well BHPs). This suggests that the GNSM does not degrade when used for unstructured versus structured models, or in cases with some amount of variation in the well controls. These interesting findings highlight the robustness of the GNSM developed in this work.

\end{document}